\documentclass[11pt]{article}

\usepackage{mathrsfs}
\usepackage{amsmath,amssymb,amsfonts,amsthm}
\usepackage{moreverb,rotating,graphics}
\usepackage[T1]{fontenc}
\usepackage[latin1]{inputenc}
\usepackage{epsfig}
\usepackage[francais,english]{babel} 
\usepackage{dsfont}
\usepackage{color}
\usepackage[normalem]{ulem}
\usepackage{caption}
\usepackage{wrapfig}

\usepackage{appendix}
\usepackage{lipsum}

\usepackage{multirow}

\newtheorem{theorem}{Theorem}

\newtheorem{remark}{Remark}

\def\tr{{\rm Tr \,}}

\def\N{{\mathbb N}}
\def\Z{{\mathbb Z}}
\def\R{{\mathbb R}}
\def\C{{\mathbb C}}
\def\S{{\mathbb S}}

\def\H{{\mathbb H}}
\def\M{{\mathbb M}}
\def\V{{\mathbb V}}

\def\P{{\mathbb P}}
\def\1{{\mathds 1}}

\def\cC{{\mathcal C}}
\def\cE{{\mathcal E}}

\def\cH{{\mathcal H}}
\def\cK{{\mathcal K}}
\def\cL{{\mathcal L}}
\def\cM{{\mathcal M}}
\def\cN{{\mathcal N}}
\def\cO{{\mathcal O}}

\def\cS{{\mathcal S}}
\def\cU{{\mathcal U}}
\def\cX{{\mathcal X}}
\def\cI{{\mathcal I}}
\def\cW{{\mathcal W}}
\def\cV{{\mathcal V}}

\def\gS{{\mathfrak S}}
\def\bL{{\mathbf L}}

\newcommand{\br}{\mathbf r}

\newcommand{\bz}{\mathbf z}
\newcommand{\bk}{\mathbf k}
\newcommand{\be}{\mathbf e}

\newcommand \dps{\displaystyle }
\newcommand{\stkout}[1]{\ifmmode\text{\sout{\ensuremath{#1}}}\else\sout{#1}\fi}

\title{A numerical study of the extended Kohn-Sham \\ ground states of atoms}

\author{Eric Canc\`es\thanks{Universit\'e Paris-Est, CERMICS, Ecole des Ponts and INRIA Paris, 6 \& 8 avenue Blaise Pascal, 77455 Marne-la-Vall\'ee Cedex 2, France, cances@cermics.enpc.fr}\and Nahia Mourad\thanks{Universit\'e Paris-Est, CERMICS, Ecole des Ponts, 6 \& 8 avenue Blaise Pascal, 77455 Marne-la-Vall\'ee Cedex 2, France, nahia.mourad@gmail.com}}

\date{\today}

\begin{document}

\selectlanguage{english}

\maketitle
\begin{abstract}
In this article, we consider the extended Kohn-Sham model for atoms subjected to cylindrically-symmetric external potentials.
The variational  approximation of the model and the construction of appropriate discretization spaces are detailed together with the algorithm to solve the discretized Kohn-Sham equations used in our code. Using this code, we compute the occupied and unoccupied energy levels of all the atoms of the first four rows of the periodic table for the reduced Hartree-Fock (rHF) and the extended Kohn-Sham X$\alpha$ models. These results allow us to test numerically the assumptions on the negative spectra of atomic rHF and Kohn-Sham Hamiltonians used in our previous theoretical works on density functional perturbation theory and pseudopotentials. Interestingly, we observe accidental degeneracies between s and d shells or between p and d shells at the Fermi level of some atoms. We also consider the case of an  atom subjected to a uniform electric-field. For various magnitudes of the electric field, we compute the response of the density of the carbon atom confined in a large ball with Dirichlet boundary conditions, and we check that, in the limit of small electric fields, the results agree with the ones obtained with first-order density functional perturbation theory.
\end{abstract}

\section{Introduction}
This article is concerned with the numerical computation of the extended Kohn-Sham ground states of atoms for the reduced Hartree-Fock (rHF, also called Hartree) and LDA (local density approximation) models~\cite{KS,PZ}. We consider the case of an isolated atom, as well as the case of an atom subjected to cylindrically symmetric external potential. We notably have in mind Stark potentials, that are potentials of the form $W(\br) = -\cE \cdot \br$ generated by a uniform electric field $\cE \neq 0$.

\medskip

We first propose a method to accurately solve the extended Kohn-Sham problem for cylindrically symmetric systems, using spherical coordinates and a separation of variables. This approach is based on the fact that, for such systems, the Kohn-Sham Hamiltonian commutes with $L_\bz$, the $\bz$-component of the angular momentum operator, $\bz$ denoting the symmetry axis of the system. We obtain in this way a family of 2D elliptic eigenvalue problems in the $r$ and $\theta$ variables, indexed by the eigenvalue $m \in \Z$ of $L_\bz$, all these problems being coupled together through the self-consistent density. To discretize the 2D eigenvalue problems, we use harmonic polynomials in $\theta$ (or in other words, spherical harmonics $Y_l^0$, which only depend on $\theta$) to discretize along the angular variable, and high-order finite element methods to discretize along the radial variable $r \in [0,L_e]$. We then apply this approach to study numerically two kinds of systems. 

\medskip

First, we provide accurate approximations of the extended Kohn-Sham ground states of all the atoms of the first four rows of the periodic table. These results allow us to test numerically the assumptions on the negative spectra of atomic rHF and Kohn-Sham LDA Hamiltonians that we used in previous theoretical works on density functional perturbation theory~\cite{CM1} and norm-conserving semilocal pseudopotentials~\cite{CM2}. We show in particular that for most atoms of the first four rows of the periodic table, the Fermi level is negative and is not an accidentally degenerate eigenvalue of the rHF Hamiltonian. We also observe that there seems to be no unoccupied orbitals with negative energies. On the other hand, for some chemical elements, the Fermi level seems to be an accidentally degenerate eigenvalue (for example the rHF 5s and 4d states of the palladium atom seem to be degenerate). For a few of them, this accidentally degenerate eigenvalue is so close to zero that our calculations do not allow us to know whether it is slightly negative or equal to zero. For instance, our simulations seem to show that the 5s and 3d states of the iron atom seem to be degenerate at the rHF level of theory, and the numerical value of their energy we obtain with our code is about $-10^{-5}$ Ha. 

\medskip

Second, we study an atom subjected to a uniform electric field (Stark effect). In this case, the system has no ground state (the Kohn-Sham energy functional is not bounded below), but density functional perturbation theory (see \cite{CM1,CM2} for a mathematical analysis) can be used to compute the polarization of the electronic cloud caused by the external electric field. The polarized electronic state is not a steady state, but a resonant state, and the smaller the electric field, the longer its life time. Another way to compute the polarization of the electronic cloud is to compute the ground state for a small enough electric field in a basis set consisting of functions decaying fast enough at infinity for the electrons to stay close to the nuclei. The Gaussian basis functions commonly used in quantum chemistry satisfy this decay property. However, it is not easy to obtain very accurate results with Gaussian basis sets, since they are not systematically improvable (over-completeness issues). Here we consider instead basis functions supported in a ball $B_{L_e}$, where $L_e$ is a numerical parameter chosen large enough to obtain accurate results and small enough to prevent electrons from escaping to infinity (for a given, small, value of the external electric field $\cE$). We study the ground state energy and density as functions of the cut-off radius $L_e$, and observe that for a given, small enough, uniform electric field, there is a plateau $[L_{e,\rm min},L_{e,\rm max}]$ on which these quantities hardly vary. For $L_e < L_{e,\rm min}$, the simulated system is too much confined, which artificially increases its energy, while for $L_e > L_{e,\rm max}$, a noticeable amount of charge accumulates at the boundary of the simulation domain, in the direction of $\cE$ (where the potential energy is very negative). On the other hand, for $L_{e,\rm min} \le L_e \le L_{e,\rm max}$, the simulation provides a fairly accurate approximation of the polarization energy and the polarized density.

\medskip

The article is organized as follows. In Section~\ref{sec:modeling}, we recall the mathematical formulation of the extended Kohn-Sham model, and some theoretical results about the rHF and LDA ground states of isolated atoms and of atoms subjected to an external cylindrically symmetric potential. In Section~\ref{sec:numerical-method}, we describe the discretization method and the algorithms used in this work to compute the extended Kohn-Sham ground states of atoms subjected to cylindrically symmetric external potentials. Some numerical results are presented in Section~\ref{sec:results}.

\section{Modeling}\label{sec:modeling}
In this article, we consider a molecular system consisting of a single nucleus of atomic charge $z\in\N^\ast$ and of $N$ electrons. For $N=z$, this system is the neutral atom with nuclear charge $z$, which we call atom $z$ for convenience.  
\subsection{Kohn-Sham models for atoms}
In the framework of the (extended) Kohn-Sham model~\cite{DreGro90}, the ground state energy of a system with one nucleus with charge $z$ and $N$ electrons is 
obtained by minimizing an energy functional of the form
\begin{equation}\label{eq:energy_KS}
E_{z,N}(\gamma) :=  \tr\left( -\frac 12 \Delta \gamma\right) -z \int_{\R^3} \frac{\rho_\gamma}{|\cdot|}  + \frac 12 D(\rho_\gamma,\rho_\gamma)+E_{\rm xc}(\rho_\gamma)
\end{equation}
over the set 
\begin{equation}\label{eq:admissibleDM}
\cK_N:=\left\{ \gamma \in \cS(L^2(\R^3)) \; | \; 0 \le \gamma \le 2, \; \tr(\gamma)=N, \; \tr(-\Delta\gamma) < \infty \right\},
\end{equation}
where $\cS(L^2(\R^3))$ is the space of the self-adjoint operators on $L^2(\R^3):=L^2(\R^3,\R)$ and $\tr(-\Delta\gamma):=\tr(|\nabla|\gamma|\nabla|)$. Note that, $\cK_N$ is a closed convex subset of the space $\gS_{1,1}$ defined by
$$
\gS_{1,1}:= \left\{T \in \gS_1 \; | \; |\nabla|T|\nabla| \in \gS_1 \right\},
$$
endowed with norm
$$
\|T\|_{\gS_{1,1}}:=\|T\|_{\gS_1}+\||\nabla \, |T|\nabla| \, \|_{\gS_1}.
$$

\medskip

The function $-\frac{z}{|.|}$ is the attraction potential induced on the electrons by the nucleus, and
$\rho_\gamma$ is the density associated with the one-body density matrix $\gamma$. For $\gamma\in\cK_N$, we have
$$
\rho_\gamma \ge 0, \quad \int_{\R^3} \rho_\gamma = N, \quad \int_{\R^3} |\nabla\sqrt{\rho_\gamma}|^2 \le \tr(-\Delta \gamma)<\infty.
$$
The last result is the Hoffmann-Ostenhof inequality~\cite{Hof77}. Therefore $\sqrt{\rho_\gamma}\in H^1(\R^3)$, and in particular, 
$\rho_{\gamma}\in L^1(\R^3)\cap L^3(\R^3)$. For $\rho\in L^{\frac 65}(\R^3)$, $D(\rho,\rho)$ is equal to  $\int_{\R^3}V^{\rm H}(\rho)\rho$, where $V^{\rm H}$ is the 
Coulomb, also called Hartree, potential generated by $\rho$: 
$$
V^{\rm H}(\rho)=\rho\star|\cdot|^{-1}.
$$
Recall that $V^{\rm H}$ can be seen as a unitary operator from the Coulomb space  ${\cal C}$ to its dual ${\cal C}'$, where
\begin{equation}\label{Coulomb_space}
\cC:=\left\{ \rho \in {\cal S}'(\R^3) \, | \, \widehat \rho \in L^1_{\rm loc}(\R^3,\C), \, |\cdot|^{-1} \widehat \rho \in L^2(\R^3,\C) \right\},\quad
(\rho_1,\rho_2)_{\cC}=4\pi\int_{\R^3}\frac{\widehat\rho_1(\bk)^\ast\widehat\rho_2(\bk)}{|\bk|^2}\,d\bk,
\end{equation}
and
\begin{equation}\label{dual_Coulomb_space}
\cC':=\left\{ v \in L^6(\R^3) \; | \; \nabla v \in (L^2(\R^3))^3 \right\}, \quad
(v_1,v_2)_{\cC'}=\frac{1}{4\pi}\int_{\R^3}\nabla v_1\nabla v_2=\frac{1}{4\pi}\int_{\R^3}|\bk|^2\widehat v_1(\bk)^\ast\widehat v_2(\bk)\, d\bk.
\end{equation}

\medskip

The term $E_{\rm xc}$ is the exchange-correlation energy. We will restrict ourselves to two kinds of Kohn-Sham models: the rHF model, for which the exchange-correlation energy is taken equal to zero 
$$E_{\rm xc}^{\rm rHF}=0,$$
and the Kohn-Sham LDA (local density approximation) model, for which the exchange-correlation energy has the form 
$$E_{\rm xc}^{\rm LDA}(\rho)=\int_{\R^3}\epsilon_{\rm xc}(\rho(r))\,dr,$$ 
where $\epsilon_{\rm xc}$ is the sum of the exchange and correlation energy densities of the homogeneous electron gas. As the function 
$\epsilon_{\rm xc}:\R_+\rightarrow\R$ is not explicitly known, it is approximated in practice by an explicit function, still denoted by $\epsilon_{\rm xc}$ for 
simplicity. We assume here that the approximate function $\epsilon_{\rm xc}$ is a $C^1$ function from $\R_+$ into $\R_-$, twice differentiable on $\R^\ast_+$ and 
obeying the following conditions
\begin{align}
&\epsilon_{\rm xc}(0)=0,\quad  \epsilon_{\rm xc}'(0)\leq 0,\label{cond:LDA1}\\
&\exists 0<\beta_-\leq \beta_+<\frac{2}{3} \quad\mbox{s.t}\quad \sup_{\rho\in\R_+}\frac{|\epsilon_{\rm xc}'(\rho)|}{\rho^{\beta_-}+
 \rho^{\beta_+}}<\infty,\label{cond:LDA2}\\
&\exists 1\leq\alpha<\frac{3}{2} \quad\mbox{s.t}\quad \limsup_{\rho\rightarrow 0_+}\frac{\epsilon_{\rm xc}(\rho)}{\rho^{\alpha}}<0\label{cond:LDA3},\\
&\exists \lambda > -1\quad \mbox{s.t} \quad \epsilon_{\rm xc}''(\rho)\underset{\rho \to 0_+}{\sim} c\rho^\lambda\label{cond:LDA4}.  
\end{align}
Note that these properties are satisfied by the exact function $\epsilon_{\rm xc}$. They are also satisfied by Slater's X$\alpha$ model for which $\epsilon_{\rm xc}(\rho)=-C_{\rm D} \rho^{\frac 13}$, where $C_{\rm D}=\frac 34\left(\frac 3\pi\right)^{\frac 13}$ is the Dirac constant. This model is used in the simulations reported in Section~\ref{sec:results}.

\medskip

\begin{remark} The minimization set $\cK_N$ defined in~(\ref{eq:admissibleDM}) is the set of {\em real spin-unpolarized} first-order reduced density matrices. We will call its elements {\em non-magnetic states}. The general (complex non-collinear spin-polarized, see e.g.~\cite{gontier}) rHF model being convex in the density matrix, and strictly convex in the density, the general rHF ground state density of a given molecular system in the absence of magnetic field, if it exists, is unique, and one of the minimizers is a non-magnetic state. Indeed, using the notation of~\cite{gontier}, if $\gamma$ is a complex non-collinear spin-polarized ground state, the non-magnetic state 
$$
\gamma_0 := \frac{1}{4} \left( \gamma^{\uparrow\uparrow}+\overline{\gamma^{\uparrow\uparrow}}+\gamma^{\downarrow\downarrow}+\overline{\gamma^{\downarrow\downarrow}}\right),
$$
where $\overline{\gamma^{\sigma,\sigma}}$ is the complex conjugate (not the adjoint) of the operator $\gamma^{\sigma,\sigma}$,
is a non-magnetic ground state. The general rHF ground state energy and density of a molecular system in the absence of magnetic field can therefore be determined by minimizing the rHF energy functional over the set $\cK_N$. The LDA model is not {\it a priori} strictly convex in the density, but it is convex over the set of complex non-collinear spin-polarized density matrices having a given density $\rho$. Therefore, the general LDA ground state energy and densities can be obtained by minimizing the LDA energy functional over the set $\cK_N$. In contrast, this argument does not apply to the local spin density approximation (LSDA) model, whose ground states are, in general, spin-polarized.
\end{remark}

\medskip

To avoid ambiguity, for any $z$ and $N$ in $\R_+^\ast$, we denote by
\begin{equation}\label{eq:min_rHF}
\cI^{\rm rHF}_{z,N} :=\inf\left\{ E_{z,N}^{\rm rHF}(\gamma),\, \gamma\in\cK_N\right\},
\end{equation}
where 
\begin{equation*}
E_{z,N}^{\rm rHF}(\gamma):= \tr\left(-\frac 12 \Delta \gamma\right)-z \int_{\R^3} \frac{\rho_\gamma}{|\cdot|}+ \frac 12 D(\rho_\gamma,\rho_\gamma),
\end{equation*}
and
\begin{equation}\label{eq:min_LDA}
\cI^{\rm LDA}_{z,N} :=\inf\left\{ E_{z,N}^{\rm LDA}(\gamma),\, \gamma\in\cK_N\right\},
\end{equation}
where
\begin{equation*}
E_{z,N}^{\rm LDA}(\gamma):= \tr\left(-\frac 12 \Delta \gamma\right)-z \int_{\R^3} \frac{\rho_\gamma}{|\cdot|}+ \frac 12 D(\rho_\gamma,\rho_\gamma)
                           +E_{\rm xc}^{\rm LDA}(\rho_\gamma).
\end{equation*}

\medskip

We recall the following two theorems which ensure the existence of ground states for neutral atoms and positive ions.

\begin{theorem}[ground state for the rHF model~\cite{CM1,Sol91}]
Let $z\in\R_+^\ast$ and $N\leq z$. Then the minimization problem (\ref{eq:min_rHF}) has a ground state $\gamma^{0,\rm rHF}_{z,N}$,  and all the ground states share the
same density $\rho^{0,\rm rHF}_{z,N}$. The mean-field Hamiltonian
$$
H^{0,\rm rHF}_{z,N} := -\frac 12 \Delta -\frac{z}{|\cdot|} + V^{\rm H}(\rho^{0,\rm rHF}_{z,N}),
$$
is a bounded below self-adjoint operator on $L^2(\R^3)$, $\sigma_{\rm ess}(H^{0,\rm rHF}_{z,N})=\R_+$, and the ground state $\gamma^{0,\rm rHF}_{z,N}$ is of the form
\begin{equation*}
\gamma^{0,\rm rHF}_{z,N}= 2\1_{(-\infty,\epsilon_{z,N,\rm F}^{0,\rm rHF})}(H^{0,\rm rHF}_{z,N} ) + \delta^{0,\rm rHF}_{z,N},
\end{equation*}
where $\epsilon_{z,N,\rm F}^{0,\rm rHF} \le 0$ is the Fermi level, $0 \le \delta^{0,\rm rHF}_{z,N}  \le 2$ and 
$\mbox{\rm Ran}(\delta^{0,\rm rHF}_{z,N} ) \subset \mbox{\rm Ker}(H^{0,\rm rHF}_{z,N} -\epsilon_{z,N,\rm F}^{0,\rm rHF})$.
If $\epsilon_{z,N,\rm F}^{0,\rm rHF}$ is negative and is not an accidentally degenerate eigenvalue of $H^{0,\rm rHF}_{z,N}$, then the non-magnetic ground state $\gamma^{0,\rm rHF}_{z,N}$ is unique.
\end{theorem}

The numerical results presented in Section~\ref{sec:numrHF} indicate that, for neutral atoms, the assumption 
\begin{center}
$\epsilon_{z,z,\rm F}^{0,\rm rHF}$ is negative and is not an accidentally degenerate eigenvalue of $H^{0,\rm rHF}_{z,z}$,
\end{center}
which guarantees the uniqueness of the non-magnetic rHF ground state density matrix, is satisfied for all the chemical elements of the first two rows of the periodic table, and for most of the elements of the third and four rows. Surprisingly, we observe accidental degeneracies at the Fermi level for Sc and Ti (4p and 3d shells), for V, Cr, Mn and Fe (5s and 3d shells), for Zr (5p and 4d shells), 
Nb and Mo (6s and 4d shells), and for Pd and Ag (5s and 4d shells). For some of these elements, the Fermi level is clearly negative, and we can conclude the following:
\begin{itemize}
\item if the Fermi level contains an s and a d shell, then the non-magnetic rHF ground state is unique;
\item if the Fermi level contains a p and a d shell, and if both shells are partially occupied (which is suggested by our numerical simulations), then the non-magnetic rHF ground state is not unique.
\end{itemize}
Indeed, in the former case, any rHF ground state is of the form 
\begin{align*}
\gamma^{0,\rm rHF}_{z,z}=2\1_{(-\infty,\epsilon_{z,z,\rm F}^{0,\rm rHF})}(H^{0,\rm rHF}_{z,z} ) & + \alpha  |\phi_{s}\rangle\langle\phi_{s}| + \sum_{m,m'=-2}^2 \beta_{m,m'} |\phi_{d,m}\rangle\langle\phi_{d,m'}| \\
& + \sum_{m=-2}^2 \gamma_m \left( |\phi_{s}\rangle\langle\phi_{d,m}| + |\phi_{d,m}\rangle\langle\phi_{s}| \right),
\end{align*}
where $\alpha \in \R$, $\beta \in \R^{5 \times 5}_{\rm sym}$ and $\gamma \in \R^{5}$ are matrices such that $0 \le \left( \begin{array}{cc} \alpha & \gamma^T \\ \gamma & \beta \end{array} \right) \le 2$ and where
$$
\phi_{s}(\br) =  f_{ns}(r), \quad \phi_{d,m}(\br) = r^2 f_{n'd}(r) \widetilde Y_2^m(\theta,\varphi).
$$
Here, the $\widetilde Y_l^m$'s are the real spherical harmonics, and $f_{ns}$ and $f_{n'd}$ are radial functions with respectively $(n-1)$ and $(n'-3)$ nodes in $(0,+\infty)$. Since all the ground state density matrices share the same density, the function 
\begin{align*}
& \alpha^2 f_{ns}(r)^2  +  \frac{\sqrt{15}}{\pi}  f_{ns}(r)f_{n'd}(r) \left( \gamma_{-2} xy +\gamma_{-1} yz + \gamma_0 \frac{2z^2-x^2-y^2}{\sqrt 3} + \gamma_1 xz  + \gamma_2 \frac{x^2-y^2}2 \right) + \frac{15}{4\pi}  f_{n'd}(r)^2 \\
 & \times  \bigg( \beta_{-2,-2} x^2y^2 + \beta_{-1,-1} y^2z^2 +\beta_{0,0}  \frac{(2z^2-x^2-y^2)^2}{\sqrt 3}   +  \beta_{1,1} x^2z^2 + \beta_{2,2} \frac{(x^2-y^2)^2}4 + 2 \beta_{-2,-1} xy^2z    \\
& \qquad +  \beta_{-2,0} \frac{xy(2z^2-x^2-y^2)}{\sqrt 3} +  2 \beta_{-2,1} x^2yz  + \beta_{-2,2} xy(x^2-y^2) +   \beta_{-1,0} \frac{yz(2z^2-x^2-y^2)}{12}+  2 \beta_{-1,1} xyz^2  \\
& \qquad   + \beta_{-1,2} yz(x^2-y^2) +  \beta_{0,1} \frac{xz(2z^2-x^2-y^2)}{\sqrt 3}  +  \beta_{0,2} \frac{(x^2-y^2)(2z^2-x^2-y^2)}{2 \sqrt 3} +   \beta_{1,2} xz(x^2-y^2)  \bigg) 
\end{align*}
where $r=(x^2+y^2+z^2)^{1/2}$, must be a function of $r$, independent of the chosen ground state density matrix. Since $f_{ns}$ has more nodes than $f_{n'd}$ (we have seen above that $n=5$ or $6$ and $n'=3$ or $4$), this implies that $\beta$ is a scalar matrix, that $\gamma=0$, and that only one value for the pair $(\alpha,\beta)$ is possible. This demonstrates the uniqueness of the non-magnetic ground state when the Fermi level is negative and contains a pair of accidentally degenerate s and d shells.

In the case when the Fermi level is negative and contains a pair of accidentally degenerate p and d shells, any non-magnetic ground state density matrix is of the form
\begin{align}
\gamma^{0,\rm rHF}_{z,z}=2\1_{(-\infty,\epsilon_{z,z,\rm F}^{0,\rm rHF})}(H^{0,\rm rHF}_{z,z} ) & + \sum_{m,m'=-1}^1 \alpha_{m,m'} |\phi_{p,m}\rangle\langle\phi_{p,m'}| + \sum_{m,m'=-2}^2 \beta_{m,m'} |\phi_{d,m}\rangle\langle\phi_{d,m'}|  \nonumber \\
& + \sum_{m=-1}^1\sum_{m'=-2}^2 \gamma_{m,m'} \left( |\phi_{p,m}\rangle\langle\phi_{d,m'}| + |\phi_{d,m'}\rangle\langle\phi_{p,m}| \right) \label{eq:admDM}
\end{align}
where $\alpha \in \R^{3 \times 3}_{\rm sym}$, $\beta \in \R^{5 \times 5}_{\rm sym}$ and $\gamma \in \R^{3 \times 5}$ are matrices such that $0 \le \left( \begin{array}{cc} \alpha & \gamma \\ \gamma^T & \beta \end{array} \right) \le 2$ and where
$$
\phi_{p,m}(\br) =  r f_{np}(r) \widetilde Y_1^m(\theta,\phi), \quad \phi_{d,m}(\br) = r^2 f_{n'd}(r) \widetilde Y_2^m(\theta,\varphi).
$$
Here, $f_{np}$ and $f_{n'd}$ are radial functions with respectively $(n-2)$ and $(n'-3)$ nodes in $(0,+\infty)$. Since all the ground state density matrices share the same density, the function 
\begin{align*}
& \frac{3}{4\pi} f_{np}(r)^2 \big( \alpha_{-1,-1} y^2+ \alpha_{0,0} z^2 + \alpha_{1,1} x^2 + 2 \alpha_{-1,0} yz + 2 \alpha_{-1,1} xy  + 2 \alpha_{0,1} xz \big) + \frac{3\sqrt 5}{2\pi}  f_{np}(r)  f_{n'd}(r)  \\ 
& \quad  \times \bigg( \gamma_{-1,-2} xy^2 + \gamma_{-1,-1} y^2z + \gamma_{-1,0} \frac{y(2z^2-x^2-y^2)}{2\sqrt 3}  + \gamma_{-1,1} xyz + \gamma_{-1,2} \frac{y(x^2-y^2)}2  \\
& \qquad\quad  \gamma_{0,-2} xyz + \gamma_{0,-1} yz^2 + \gamma_{0,0} \frac{z(2z^2-x^2-y^2)}{2\sqrt 3}  + \gamma_{0,1} xz^2 + \gamma_{0,2} \frac{z(x^2-y^2)}2 \\
& \qquad\quad  \gamma_{1,-2} x^2y + \gamma_{1,-1} xyz + \gamma_{1,0} \frac{x(2z^2-x^2-y^2)}{2\sqrt 3}  + \gamma_{1,1} x^2z + \gamma_{1,2} \frac{x(x^2-y^2)}2 
\bigg) + \frac{15}{4\pi}  f_{n'd}(r)^2 \\
 & \times  \bigg( \beta_{-2,-2} x^2y^2 + \beta_{-1,-1} y^2z^2 +\beta_{0,0}  \frac{(2z^2-x^2-y^2)^2}{12}   +  \beta_{1,1} x^2z^2 + \beta_{2,2} \frac{(x^2-y^2)^2}4 + 2 \beta_{-2,-1} xy^2z    \\
& \qquad +  \beta_{-2,0} \frac{xy(2z^2-x^2-y^2)}{\sqrt 3} +  2 \beta_{-2,1} x^2yz  + \beta_{-2,2} xy(x^2-y^2) +   \beta_{-1,0} \frac{yz(2z^2-x^2-y^2)}{\sqrt 3}+  2 \beta_{-1,1} xyz^2  \\
& \qquad   + \beta_{-1,2} yz(x^2-y^2) +  \beta_{0,1} \frac{xz(2z^2-x^2-y^2)}{\sqrt 3}  +  \beta_{0,2} \frac{(x^2-y^2)(2z^2-x^2-y^2)}{2 \sqrt 3} +   \beta_{1,2} xz(x^2-y^2)  \bigg) 
\end{align*}
where $r=(x^2+y^2+z^2)^{1/2}$, must be a function of $r$, independent of the chosen ground state density matrix. Since $f_{np}$ has more nodes than $f_{n'd}$, this implies that $\alpha$ and $\beta$ are scalar matrices and that, for $\alpha$ and $\beta$ given, the function 
\begin{align*}
& \gamma_{-1,-2} xy^2 + \gamma_{-1,-1} y^2z + \gamma_{-1,0} \frac{y(2z^2-x^2-y^2)}{2\sqrt 3}  + \gamma_{-1,1} xyz + \gamma_{-1,2} \frac{y(x^2-y^2)}2  \\
& \qquad\quad  \gamma_{0,-2} xyz + \gamma_{0,-1} yz^2 + \gamma_{0,0} \frac{z(2z^2-x^2-y^2)}{2\sqrt 3}  + \gamma_{0,1} xz^2 + \gamma_{0,2} \frac{z(x^2-y^2)}2 \\
& \qquad\quad  \gamma_{1,-2} x^2y + \gamma_{1,-1} xyz + \gamma_{1,0} \frac{x(2z^2-x^2-y^2)}{2\sqrt 3}  + \gamma_{1,1} x^2z + \gamma_{1,2} \frac{x(x^2-y^2)}2 
\end{align*}
is a given function of $r$. The vector spaces of homogeneous polynomials in $x,y,z$ of total degree equal to $3$ is of dimension 10, and the matrix $\gamma$ has 15 independent entries. Provided $\alpha$ and $\beta$ are not equal to $0$ (which is suggested by our numerical simulations), an infinity of density matrices of the form (\ref{eq:admDM}) satisfy the rHF equations, and are therefore admissible non-magnetic ground states.

For other chemical elements, such as iron ($z=26)$, the Fermi level is so close to zero that the numerical accuracy of our numerical method does not allow us to know whether it is slightly negative or equal to zero.

\begin{theorem}[ground state for the LDA model~\cite{AnaCan09}]
Let $z\in\R_+^\ast$ and $N\leq z$. Suppose that (\ref{cond:LDA1})-(\ref{cond:LDA3}) hold. Then the minimization problem (\ref{eq:min_LDA}) has a ground state 
$\gamma^{0,\rm LDA}_{z,N}$. In addition, $\gamma^{0,\rm LDA}_{z,N}$ satisfies the self-consistent field equation
\begin{equation} \label{eq:ground_state_LDA}
\gamma^{0,\rm LDA}_{z,N} = 2\1_{(-\infty,\epsilon_{z,N,\rm F}^{0,\rm LDA})}(H^{0,\rm LDA}_{z,N}) + \delta^{0,\rm LDA}_{z,N},
\end{equation}
where $\epsilon_{z,N,\rm F}^{0,\rm LDA} \le 0$ is the Fermi level, $0 \le \delta^{0,\rm LDA}_{z,N}\le 2$, 
$\mbox{\rm Ran}(\delta^{0,\rm LDA}_{z,N}) \subset \mbox{\rm Ker}(H^{0,\rm LDA}_{z,N}-\epsilon_{z,N,\rm F}^{0,\rm LDA} )$ and the mean-field Hamiltonian
$$
H^{0,\rm LDA}_{z,N} := -\frac 12 \Delta -\frac{z}{|\cdot|} + V^{\rm H}(\rho^{0,\rm LDA}_{z,N})+v_{\rm xc}(\rho^{0,\rm LDA}_{z,N}),
$$
where $\rho^{0,\rm LDA}_{z,N}=\rho_{\gamma^{0,\rm LDA}_{z,N}}$ and $v_{\rm xc}(\rho)=\frac{d\,\epsilon_{\rm xc }}{d\, \rho}(\rho)$, is a bounded below self-adjoint 
operator on $L^2(\R^3)$ and $\sigma_{\rm ess}(H^{0,\rm LDA}_{z,N})=\R_+$.
\end{theorem}

\subsection{Density functional perturbation theory}\label{sec:perturbation}

We now examine the response of the ground state density matrix when an additional external potential $\beta W$ is turned on. The energy functional to be minimized over 
$\cK_N$ now reads
\begin{equation}\label{eq:pert_energy}
 \widetilde E_{z,N}^{\rm rHF/LDA}(\gamma,\beta W):=E_{z,N}^{\rm rHF/LDA}(\gamma)+\int_{\R^3}\beta W\rho_{\gamma},
\end{equation}
and is well-defined for any $\gamma\in\cK_N$, $W\in\cC'$ and $\beta\in\R$. The parameter $\beta$ is called the coupling constant in quantum mechanics. Denote by

\begin{equation}\label{eq:min_pert}
\widetilde\cI_{z,N}^{\rm rHF/LDA}(\beta W) :=\inf\left\{\widetilde E_{z,N}^{\rm rHF/LDA}(\gamma,\beta W), \, \gamma\in\cK_N\right\}.
\end{equation}

\medskip

The following theorem insures the existence of a perturbed ground state density matrix for perturbation potentials in $\cC'$. 

\begin{theorem}[existence of a perturbed minimizer~\cite{CM1}] \label{Th:perturbC'}
Let $z\in\R_+^\ast$, $N\leq z$ and $W\in\cal C'$. Assume that the Fermi level $\epsilon_{z,N,\rm F}^{0,\rm rHF}$ is negative and is not an accidentally degenerate eigenvalue of $H^{0,\rm rHF}_{z,N}$. Then the non-magnetic unperturbed rHF ground state, that is the minimizer of (\ref{eq:min_rHF}), is unique, and the perturbed problem (\ref{eq:min_pert}) has a unique non-magnetic ground state $\gamma_{z,N,\beta W}^{\rm rHF}$, for $\beta\in\R$ small enough. The Hamiltonian 
\begin{equation}\label{eq:H^W}
H_{z,N,\beta W}^{\rm rHF}=-\frac 12\Delta-\frac{z}{|\cdot|}+V^{\rm H}(\rho_{z,N,\beta W}^{\rm rHF})+\beta W,
\end{equation}
where $\rho_{z,N,\beta W}^{\rm rHF}=\rho_{\gamma_{z,N,\beta W}^{\rm rHF}}$, is a bounded below self-adjoint operator on $L^2(\R^3)$ with form domain $H^1(\R^3)$ and
$\sigma_{\rm ess}(H^{0,\rm rHF}_{z,N,\beta W})=\R_+$. Moreover, 
$\gamma_{z,N,\beta W}^{\rm rHF}$ and $\rho_{z,N,\beta W}^{\rm rHF}$ are analytic in $\beta$, that is
$$
\gamma_{z,N,\beta W}^{\rm rHF}=\sum_{k\geq 0}\beta^k\gamma_{z,N,W}^{(k),\rm rHF}\quad\mbox{and}\quad 
\rho_{z,N,\beta W}^{\rm rHF}=\sum_{k\geq 0}\beta^k\rho_{z,N,W}^{(k),\rm rHF},
$$
the above series being normally convergent in $\gS_{1,1}$ and $\cC$ respectively. 
\end{theorem}
In the sequel, we will refer to  $\gamma_{z,N,W}^{(k)}$ as the $k$-th order perturbation of the density matrix.

\medskip

Although we focus here on non-magnetic states, it is convenient to consider $H_{z,N}^{0,\rm rHF}$ as an operator on $L^2(\R^3,\C)$ in order to expand the angular part of the atomic orbitals on the usual complex spherical harmonics. It would of course have been possible to avoid considering complex wave functions by expanding on real spherical harmonics. However, we have chosen to work with complex wave function to prepare the ground for future works on magnetic systems.

The unperturbed Hamiltonian $H_{z,N}^{0,\rm rHF}$ is a self-adjoint operator on $L^2(\R^3,\C)$ invariant with respect to rotations around the nucleus (assumed located at 
the origin). This operator is therefore block-diagonal in the decomposition of $L^2(\R^3,\C)$ as the direct sum of the pairwise orthogonal subspaces 
$\cH_l:=\mbox{Ker}(\bL^2-l(l+1))$:
\begin{equation*} \label{eq:dec_L2}
L^2(\R^3,\C) = \bigoplus_{l \in \N} \cH_l,
\end{equation*}
where $\bL= \br \times (-i\nabla)$ is  the angular momentum operator. Since we are going to consider perturbation potentials which are not spherically
symmetric, but only cylindrically symmetric, or in other words independent of the azimuthal angle $\varphi$ in spherical coordinates, the $\cH_l$'s are no longer invariant subspaces of the perturbed 
Hamiltonians. The appropriate decomposition of $L^2(\R^3,\C)$ into invariant subspaces for Hamiltonians $H_{z,N,\beta W}^{\rm rHF}$ with $W$  cylindrically symmetric,
is the following: for $m\in\Z$, we set 
$$
{\cal H}^m:=\mbox{Ker}(L_{\bz}-m),
$$
where $L_\bz$ is the $\bz$-component of the angular momentum operator $\bL$ ($L_{\bz}=\bL.\be_{\bz}$). 

Note that
$$
\forall l \in \N, \quad \cH_l=\left\{\phi\in L^2(\R^3,\C),\quad \mbox{s.t}\quad \phi(r,\theta,\varphi)=\dps\sum_{-l\leq m\leq l}R^m(r)Y_l^m(\theta,\varphi)\right\},
$$
and
$$
\forall m \in \Z, \quad {\cal H}^m=\left\{\phi\in L^2(\R^3,\C),\quad \mbox{s.t}\quad \phi(r,\theta,\varphi)=\dps\sum_{l\geq |m|}R_l(r)Y_l^m(\theta,\varphi)\right\},
$$
where $Y_l^m$ are the spherical harmonics, i.e. the joint eigenfunctions of $\Delta_S$, the Laplace-Beltrami operator on the unit sphere $\S^2$ of $\R^3$, and $\cL_{\bz}$, the generator of rotations about the azimuthal axis of $\S^2$. More precisely, we have
\begin{equation*}
 -\Delta_S Y_l^m=l(l+1)Y_l^m\quad\text{and}\quad \cL_{\bz}Y_l^m=mY_l^m,
\end{equation*}
where, in spherical coordinates,
$$
\Delta_S=\frac{1}{\sin\theta}\frac{\partial}{\partial\theta}\left(\sin\theta\frac{\partial}{\partial\theta}\right)
           +\frac{1}{\sin^2\theta}\frac{\partial^2}{\partial\varphi^2}\quad\mbox{and}\quad \cL_{\bz}=-i\frac{\partial}{\partial\varphi}.
$$
These functions are orthonormal, in the following sense:
\begin{equation}\label{eq:orthoYlm}
 \int_{\S^2}Y_l^m(Y_{l'}^{m'})^*=\int_{\theta=0}^\pi\int_{\varphi=0}^{2\pi}Y_l^m(\theta,\varphi)\left(Y_{l'}^{m'}(\theta,\varphi)\right)^* \sin\theta\,d\theta\,d\varphi
             =\delta_{ll'}\delta_{mm'},
\end{equation}
where $\delta_{ij}$ is the Kronecker symbol and $(Y_l^m)^*=(-1)^mY_l^{-m}$ is the complex conjugate of $Y_l^m$. 

We also define
$${\cal V}^m:={\cal H}^m\cap  H^1(\R^3,\C),$$
so that $L^2(\R^3,\C)$ and $H^1(\R^3,\C)$ are decomposed as the following direct sums: 
\begin{equation}\label{eq:space-decomposition}
L^2(\R^3,\C)=\dps\bigoplus_{m\in \Z}{\cal H}^m\quad\mbox{and}\quad H^1(\R^3,\C)=\dps\bigoplus_{m\in \Z}{\cal V}^m,
\end{equation}
each $\cH^m$ being $H_{z,N,\beta W}^{\rm rHF}$-stable (in the sense of unbounded operators) for $W$ cylindrically symmetric. This is due to the fact that, for $W$ cylindrically symmetric, the operator $H_{z,N,\beta W}^{\rm rHF}$ commutes with $L_{\bz}$. Note that $\sigma(H_{z,N,\beta W}^{\rm rHF})=\overline{\underset{m\in \Z}{\cup}\sigma\left(H_{z,N,\beta W}^{\rm rHF}|_{{\cal H}^m}\right)}$. Same arguments hold true for $H_{z,N,\beta W}^{\rm LDA}$ under the assumption that the ground state density $\rho_{z,N,\beta W}^{0,\rm LDA}$ is cylindrically symmetric (which is the case whenever it is unique).

\medskip

We are interested in the Stark potential 
\begin{equation}\label{eq:WStark}
W_{\rm Stark}(\br)=- e_{\bz} \cdot\br,
\end{equation}
 which does not belong to $\cC'$, and thus does not fall into the scope of Theorem~\ref{Th:perturbC'}. We therefore introduce the classes of perturbation potentials
$$
\cW_s:=\left\{W\in \cH^0_{\rm loc} \; | \; \int_{\R^3}\frac{|W(\br)|^2}{(1+|\br|^2)^s} \, d\br<\infty \right\},
$$
where $\cH^0_{\rm loc}:=\cH^0\cap L^2_{\rm loc}(\R^3)$, which contain the Stark potential $W_{\rm Stark}$ whenever $s > 5/2$. For $W\in\cW_s\setminus \cC'$, the energy functional (\ref{eq:pert_energy}) is not necessarily bounded below on $\cK_N$ for $\beta\neq 0$. Thus the solution of (\ref{eq:min_pert}) may not exist. This is the case for the Stark potential $W_{\rm Stark}$. However, the $k$-th order perturbation of the ground state may exist, as this is the case when the linear Schr\"odinger operator of the hydrogen atom is perturbed by the Stark potential $W_{\rm Stark}$ (see e.g~\cite{ReeSim78}). The following theorem ensures the existence of the first order perturbation of the density matrix.  
  
\begin{theorem}[first order density functional perturbation theory~\cite{CM2}]\label{th:stark_effect}
Let $z\in\R_+^\ast$, $0<N\leq z$, such that $\epsilon_{z,N,\rm F}^{0,\rm rHF}$ is negative\footnote{Note that, $\epsilon_{z,N,\rm F}^{0,\rm rHF}<0$ whenever $0<N< z$ (see e.g.~\cite{Sol91}).} and is not an accidentally degenerate eigenvalue of $H_{z,N}^{0,\rm rHF}$, $s \in \R$ and $W\in\cW_s$. In the rHF framework, the first order perturbation of the density matrix $\gamma_{z,N,W}^{(1),\rm rHF}$ is well defined in $\gS_{1,1}$.
\end{theorem}
Note that assumption (\ref{cond:LDA4}) is used to establish the existence and uniqueness of the first order perturbation of the density matrix $\gamma_{z,N,W}^{(1),\rm LDA}$ in $\gS_{1,1}$.

\section{Numerical method}\label{sec:numerical-method}

In this section, we present the discretization method and the algorithms we used to calculate numerically the ground state density matrices for (\ref{eq:min_rHF}), 
(\ref{eq:min_LDA}) and (\ref{eq:min_pert}) for cylindrically symmetric perturbation potentials $W$, together with the ground state energy and the lowest eigenvalues of the 
associated Kohn-Sham operator. From now on, we make the assumption that the ground state density of (\ref{eq:min_pert}), if it exists, is cylindrically symmetric which is 
always the case for the rHF model. Using spherical coordinates, we can write
$$
W(r,\theta)=\dps\sum_{l=0}^{+\infty} W_l(r)Y_l^0(\theta)\in\cH^0
$$
(since $Y_l^0$ is independent of $\varphi$, we use the notation $Y_l^0(\theta)$ instead of $Y_l^0(\theta,\varphi)$). As the  ground state density  
$\rho_{z,N,\beta W}$ is assumed to be cylindrically symmetric as well, one has 
$$
\rho_{z,N,\beta W}(r,\theta)=\sum_{l=0}^{+\infty}\rho_{z,N,\beta W,l}(r)Y_l^0(\theta).
$$

The Hartree and the exchange-correlation potentials also have the same symmetry. For $\rho\in L^1(\R^3)\cap L^3(\R^3)\cap\cH^0$, we have
$$
V^{\rm H}(\rho)(r,\theta)=\sum_{l=0}^{+\infty} V^{\rm H}_{\rho_l}(r)Y_l^0(\theta),\quad\mbox{and}\quad 
v_{\rm xc}(\rho)(r,\theta)=\sum_{l=0}^{+\infty} (v_{\rho}^{\rm xc})_l(r)Y_l^0(\theta),
$$
where, for each $l\geq 0$, $ V^{\rm H}_{\rho_l}(r)$ solves the following differential equation
\begin{equation*}
 -\frac{1}{r}\frac{d^2}{dr^2}(rV^{\rm H}_{\rho_l})+\frac{l(l+1)}{r^2}V^{\rm H}_{\rho_l}=4\pi\rho_l
 \end{equation*}
 with boundary conditions 
 $$
 \lim_{r \to 0^+} rV^{\rm H}_{\rho_l}(r)=0 \quad \mbox{and} \quad \lim_{r \to +\infty} rV^{\rm H}_{\rho_l}(r) = \left( 4\pi \int_0^{+\infty} r^2\rho_0(r) \, dr\right)\delta_{l0}, 
 $$
while $(v_\rho^{\rm xc})_l$ can be computed by projection on the spherical harmonics $Y_l^0$:
\begin{equation*}
 (v_\rho^{\rm xc})_l(r)=2\pi\int_{0}^\pi v_{\rm xc}(\rho)(r,\theta)Y_l^0(\theta)\sin\theta d\theta.
\end{equation*}

\subsection{Discretisation of the Kohn-Sham model}\label{sec:discretization}
Recall that for $W\in\cW_s$ and $\beta\ne 0$, the energy functional defined by (\ref{eq:pert_energy}) is not necessarily bounded below on $\cK_N$, which implies in 
particular that (\ref{eq:min_pert}) may have no ground state. Nevertheless, one can compute approximations of (\ref{eq:min_pert}) in finite-dimensional spaces, 
provided that the basis functions decay fast enough at infinity. Let $N_h\in \N^\ast$ and 
$m_h\geq m_z^*:=\max\{m|\, \exists k>0;\, \epsilon_{m,k}^0\leq \epsilon_{z,N,\rm F}^0\}$, and  
let $\{\cX_i\}_{1\leq i\leq N_h}\in \left(H^1_0(0,+\infty)\right)^{N_h}$ be a free family of real-valued basis functions. We then introduce the finite-dimensional spaces
\begin{equation}\label{eq:Vmh}
\cV^{m,h}:=\cV^m\cap\mbox{span}_\R\left(\frac{\cX_i(r)}{r}Y_l^m(\theta,\phi)\right)_{\underset{|m|\leq l\leq m_h}{1\leq i\leq N_h}}\subset H^1(\R^3,\C)
\end{equation}
and
\begin{equation}\label{eq:Xh}
\cX^h={\mbox{span}_\R(\cX_1,\cdots,\cX_{N_h})}\subset H^1_0(0,+\infty),
\end{equation}
and the set
\begin{equation*}\label{def:approx-K_N}
 \cK_{N,h}:=\left\{\gamma\in\cK_N|\quad \gamma=\sum_{m=-m_h}^{m_h}\gamma^m,\quad \gamma^m\in\cS(\cH^m),\quad\mbox{and}\quad\mbox{Ran}(\gamma^m)\subset\cV^{m,h}\right\} \subset \cK_N.
\end{equation*}
Note that since our goal is to compute non-magnetic ground states, we are allowed to limit ourselves to real linear combinations in (\ref{eq:Vmh}) and (\ref{eq:Xh}).

\subsubsection{Variational approximation}
A variational approximation of (\ref{eq:min_pert}) is obtained by minimizing the energy functional (\ref{eq:pert_energy}) over the approximation set $\cK_{N,h}$: 
\begin{equation}\label{eq:min_pert_approx}
\widetilde\cI_{z,N,h}^{\rm rHF/LDA}(\beta W) :=\inf\left\{\widetilde E_{z,N}^{\rm rHF/LDA}(\gamma_h,\beta W), \; \gamma_h\in\cK_{N,h}\right\}.
\end{equation}
Any $\gamma_h\in\cK_{N,h}$ can be written as
\begin{equation}\label{eq:gammah}
\gamma_h=\dps\sum_{\underset{1\leq k\leq (m_h-|m|+1)  N_h}{-m_h\leq m\leq m_h}}n_{m,k}|\Phi_{m,k,h}\rangle\langle\Phi_{m,k,h}|,
\end{equation}
with
$$
\Phi_{m,k,h}\in \cV^{m,h},\quad \int_{\R^3}\Phi_{m,k,h}\Phi_{m,k',h}^\ast=\delta_{kk'},\quad \Phi_{-m,k,h} = (-1)^m \Phi_{m,k,h}^\ast, 
$$
$$
 0\leq n_{m,k}=n_{-m,k} \leq 2, 
\quad \dps\sum_{\underset{1\leq k\leq (m_h-|m|+1) N_h}{-m_h\leq m\leq m_h}}n_{m,k}=N.
$$
The functions $\Phi_{m,k,h}$ being in $\cV^{m,h}$, they are of the form 
\begin{equation}\label{eq:phih}
\Phi_{m,k,h}(r,\theta,\varphi)=\sum_{l=|m|}^{m_h} \frac{u_{l}^{m,k,h}(r)}{r} Y_{l}^{m}(\theta,\varphi),
\end{equation}
where for each $-m_h \leq m \leq m_h$, $1\leq k\leq (m_h-|m|+1) N_h$ and $|m|\leq l \leq m_h$, $u_{l}^{m,k,h} \in \cX^h$. Note that $u_{l}^{-m,k,h} =u_{l}^{m,k,h}$.
Expanding the functions $u_{l}^{m,k,h}$  in the basis $(\cX_i)_{1\leq i\leq N_h}$ as
\begin{equation}
\label{eq:expandul}
u_{l}^{m,k,h}(r)=\sum_{i=1}^{N_h} U_{i,l}^{m,k} \cX_i(r),
\end{equation}
and gathering the coefficients $U_{i,l}^{m,k}$ for fixed $m$ and $k$ in a rectangular matrix $U^{m,k} \in\R^{N_h \times (m_h-|m|+1)}$, any $\gamma_h \in \cK_{N,h}$ can be represented via (\ref{eq:gammah})-(\ref{eq:expandul}) by at least one element of the set
\begin{equation}\label{eq:desgh}
\cM_{N,h}:= \cU_{h} \times \cN_{N,h}, 
\end{equation}
where
$$ \!\!\!\!\!\!\!\!\!\!\!\!\!
\cU_{h}:=\left\{(U^{m,k})_{\underset{1\leq k\leq (m_h-|m|+1) N_h}{-m_h\leq m\leq m_h}} \; | \;  U^{m,k} = U^{-m,k} \in \R^{N_h \times (m_h-|m|+1)},  \;  \tr( [U^{m,k}]^T M_0 U^{m,k'}) = \delta_{kk'}  \right\},
$$
and 
$$ \!\!\!\!\!\!\!\!\!\!\!\!\!\!\!\!
\cN_{N,h}:=\left\{(n_{m,k})_{\underset{1\leq k\leq (m_h-|m|+1) N_h}{-m_h\leq m\leq m_h}}, \;  0 \le n_{m,k}=n_{-m,k} \le 2, \; \sum_{\underset{1\leq k\leq (m_h-|m|+1) N_h}{-m_h\leq m\leq m_h}}n_{m,k}=N \right\}.
$$
The matrix $M_0$ appearing in the definition of $\cU_{h}$ is the mass matrix defined by
$$
[M_0]_{ij} = \int_0^{+\infty} \cX_i \cX_j,
$$
and the constraints $\tr( [U^{m,k}]^T M_0 U^{m,k'}) = \delta_{kk'}$ come from the fact that
\begin{align*}
\int_{\R^3}\Phi_{m,k,h}\Phi_{m,k',h}^\ast &= \int_0^{+\infty} \int_{\S^2} \left( \sum_{l=|m|}^{m_h}  \sum_{i=1}^{N_h} U_{i,l}^{m,k} \frac{\cX_i(r)}{r} Y_{l}^{m}(\sigma) \right)
 \left( \sum_{l'=|m|}^{m_h} \sum_{i=1}^{N_h} U_{j,l'}^{m,k'} \frac{\cX_j(r)}{r} Y_{l'}^{m}(\sigma)^\ast \right) \, r^2 \, d\sigma \, dr \\
 &=  \sum_{l=|m|}^{m_h}  \sum_{i,j=1}^{N_h} U_{i,l}^{m,k} [M_0]_{ij} U_{j,l}^{m,k'} = \tr( [U^{m,k}]^T M_0 U^{m,k'}). 
 \end{align*}

\begin{remark}
An interesting observation is that, if there is no accidental degeneracy in the set of the occupied energy levels of $H_{z,N}^{0,\rm rHF/LDA}$, and if the occupied orbitals are well enough approximated in the space $\cV^{m,h}$, then the approximate ground state density matrix $\gamma^{0,\rm rHF/LDA}_{z,N,h}$ has a unique representation of the form (\ref{eq:gammah})-(\ref{eq:expandul}), up to the signs and the numbering of the functions $u_l^{m,k,h}$, that is up to the signs and numbering of the column vectors of the matrices $U^{m,k}$. By continuity, this uniqueness of the representation will survive if a small-enough cylindrically-symmetric perturbation is turned on. This is the reason why this representation is well-suited to our study.
\end{remark}

Let us now express each component of the energy functional $\widetilde E_{z,N}^{\rm rHF,LDA}(\gamma_h,\beta W)$ using the representation (\ref{eq:gammah})-(\ref{eq:expandul})  of the elements of $\cK_{N,h}$. For this purpose, we introduce the $N_h\times N_h$ real symmetric matrices $A$ and $M_n$, $n=-2,-1,0,1$ with entries
\begin{eqnarray}\label{matrices}
A_{ij}=\int_0^{+\infty} \cX_i'\cX_j' \quad \mbox{and} \quad [M_n]_{ij}=\int_0^{+\infty}  r^n \cX_i(r) \cX_j(r) \, dr.
\end{eqnarray}
The weighted mass matrices $M_{-2}$ and $M_{-1}$ are well-defined in view of the Hardy inequality
$$
\forall u \in H^1_0(0,+\infty), \quad \int_0^{+\infty} \frac{u^2(r)}{r^2} \, dr \le 4 \pi \int_0^{+\infty} |u'|^2.
$$
We assume from now on that the basis functions $\cX_i$ decay fast enough at infinity for the weighted mass matrix $M_1$ to be well-defined.

\medskip

In the representation (\ref{eq:gammah})-(\ref{eq:expandul}), the kinetic energy is equal to
\begin{equation}\label{eq:kin}
\frac 12\tr(-\Delta \gamma_h)=\frac 12\dps\sum_{\underset{1\leq k\leq (m_h-|m|+1)\times N_h}{-m_h\leq m\leq m_h}} n_{m,k}
                   \dps  \left( \tr\left( [U^{m,k}]^T A U^{m,k} \right) + \tr\left( D_m [U^{m,k}]^T M_{-2} U^{m,k} \right)   \right),
\end{equation}
where $D_m \in \R^{(m_h-|m|+1)\times (m_h-|m|+1)}$ is the diagonal matrix defined by
\begin{equation}\label{eq:matDm}
D_m=\mbox{diag}(|m|(|m|+1), \cdots, m_h(m_h+1)).
\end{equation}
All the other terms in the energy functional depending on the density 
\begin{equation}\label{eq:rhoh}
\rho_h:=\rho_{\gamma_h}=\dps\sum_{\underset{1\leq k\leq (m_h-|m|+1)  N_h}{-m_h\leq m\leq m_h}}n_{m,k}|\Phi_{m,k,h}|^2,
\end{equation}
we first need to express this quantity as a function of the matrices $U^{m,k}$ and the occupation numbers $n_{m,k}$.  As the function $\rho_h$ is in $\cH^0$, we have
\begin{equation}\label{eq:rhoh2}
\rho_h(r,\theta)=\sum_{l=0}^{2m_h}\rho_l^h(r)Y_{l}^0(\theta).
\end{equation}
Inserting (\ref{eq:phih}) in (\ref{eq:rhoh}), we get  
\begin{equation}\label{eq:rhohexp}
\rho_h(r,\theta)=\dps\sum_{\underset{1\leq k\leq (m_h-|m|+1)N_h}{-m_h\leq m\leq m_h}}n_{m,k}
                \left|\sum_{l=|m|}^{m_h}\frac{u_{l}^{m,k,h}(r)}r Y_{l}^{m}(\theta,\varphi)\right|^2.
\end{equation}
We recall the following equality~\cite{Rose57}
\begin{equation}\label{eq:prodYlm}
Y_{l_1}^m(Y_{l_2}^{m})^\ast=(-1)^{m}Y_{l_1}^mY_{l_2}^{-m}=\dps\sum_{l_3=|l_1-l_2|}^{l_1+l_2} c^m_{l_1,l_2,l_3}Y_{l_3}^0,
\end{equation}
with
$$
 c^m_{l_1,l_2,l_3}= (-1)^{m}\sqrt{\frac{(2l_1+1)(2l_2+1)(2l_3+1)}{4\pi}}\left(\begin{matrix} l_1 & l_2 & l_3\\ m & -m& 0 \end{matrix}\right)
\left(\begin{matrix} l_1 & l_2 & l_3\\ 0 & 0& 0 \end{matrix}\right),
$$
where $\dps \left(\begin{matrix} l_1 & l_2 & l_3\\ m_1 & m_2& m_3 \end{matrix}\right)$ denote the Wigner 3j-symbols. Inserting the expansion (\ref{eq:expandul}) in (\ref{eq:rhohexp}) and using  (\ref{eq:prodYlm}) and the fact that
$$
\left(\begin{matrix} l_1 & l_2 & l_3\\ m_1 & m_2& m_3 \end{matrix}\right)=0 \quad \mbox{unless} \quad |l_1-l_2| \le l_3 \le l_1+l_2,
$$
we obtain 
$$
\rho_h(r,\theta)=\dps\sum_{l=0}^{2m_h}\left[\dps\sum_{i,j=1}^{N_h}\left(\dps\sum_{\underset{1\leq k\leq (m_h-|m|+1)\times N_h}{-m_h\leq m\leq m_h}}n_{m,k}
               \sum_{l',l''=|m|}^{m_h}c_{l',l'',l}^mU_{i,l'}^{m,k}U_{j,l''}^{m,k}\right) \frac{\cX_i(r)}{r}\frac{\cX_j(r)}{r}\right]Y_l^0(\theta),
$$
from which we conclude that
$$
\rho_l^h(r)= \dps\sum_{i,j=1}^{N_h}\left(\dps\sum_{\underset{1\leq k\leq (m_h-|m|+1)\times N_h}{-m_h\leq m\leq m_h}}n_{m,k}
             \sum_{l',l''=|m|}^{m_h}c_{l',l'',l}^mU_{i,l'}^{m,k}U_{j,l''}^{m,k}\right)\frac{\cX_i(r)}{r}\frac{\cX_j(r)}{r}.
$$
For $0\leq l\leq 2m_h$, we introduce the matrix $R_l \in \R^{N_h\times N_h}$ defined by
\begin{equation}\label{eq:defRl}
R_l := \dps\sum_{\underset{1\leq k\leq (m_h-|m|+1)\times N_h}{-m_h\leq m\leq m_h}}n_{m,k} U^{m,k} C^{l,m} [U^{m,k}]^T
\end{equation}
where $C^{l,m} \in \R^{(m_h-|m|+1) \times (m_h-|m|+1)}$ is the symmetric matrix\footnote{The symmetry of the matrix $C^{lm}$ comes from the following symmetry properties of the 3j-symbols:
$$
\left(\begin{matrix} l_1 & l_2 & l_3\\ m_1 & m_2 & m_3 \end{matrix}\right) = (-1)^{l_1+l_2+l_3} \left(\begin{matrix} l_2 & l_1 & l_3\\ m_2 & m_1& m_3 \end{matrix}\right) = (-1)^{l_1+l_2+l_3} \left(\begin{matrix} l_2 & l_1 & l_3\\ -m_2 & -m_1& -m_3 \end{matrix}\right).
$$
} defined by
\begin{equation}\label{eq:matrixClm}
\forall |m| \le l \le 2 m_h, \quad C^{l,m}_{l',l''}=  \sqrt{4\pi} \, c_{l',l'',l}^m,
\end{equation}
so that
\begin{equation}\label{rho}
 \rho_h(r,\theta)= \frac{1}{\sqrt{4\pi}} \dps\sum_{l=0}^{2m_h}\dps\sum_{i,j=1}^{N_h} [R_l]_{i,j} \frac{\cX_i(r)}{r}\frac{\cX_j(r)}{r} Y_l^0(\theta).
\end{equation}
Note that $C^{0,m}$ is the identity matrix, so that
$$
R_0 = \dps\sum_{\underset{1\leq k\leq (m_h-|m|+1)\times N_h}{-m_h\leq m\leq m_h}}n_{m,k} U^{m,k}  [U^{m,k}]^T
$$
and
$$
\tr(M_0R_0) = \sum_{\underset{1\leq k\leq (m_h-|m|+1)\times N_h}{-m_h\leq m\leq m_h}}n_{m,k} \tr(M_0U^{m,k}  [U^{m,k}]^T) =  \sum_{\underset{1\leq k\leq (m_h-|m|+1)\times N_h}{-m_h\leq m\leq m_h}}n_{m,k} = N,
$$
and that $C^{1,m}$ is a symmetric tridiagonal matrix whose diagonal elements all are equal to zero.

\medskip

The Coulomb attraction energy between the nucleus and the electrons then is equal to
\begin{align*}\label{coulomb-attraction}
-z \int_{\R^3} \frac{\rho_h}{|\cdot|} &= - z \int_0^{+\infty} \int_{\S^2} \frac{1}{r}  \, \left( \frac{1}{\sqrt{4\pi}} \dps\sum_{l=0}^{2m_h}\sum_{i,j=1}^{N_h} [R_l]_{i,j} \frac{\cX_i(r)}{r}\frac{\cX_j(r)}{r} Y_l^0(\sigma) \right)  r^2 \, dr \, d\sigma \\ &= - z   \int_0^{+\infty} \int_{\S^2} \frac{1}{r}  \, \left(  \dps\sum_{l=0}^{2m_h}\sum_{i,j=1}^{N_h} [R_l]_{i,j} \frac{\cX_i(r)}{r}\frac{\cX_j(r)}{r} Y_l^0(\sigma) \right) Y_0^0(\sigma)^\ast \,  r^2 \, dr \, d\sigma \\ &=-z \dps\sum_{i,j=1}^{N_h}[R_0]_{i,j} [M_{-1}]_{ij} = -z  \tr(M_{-1}R_0),
\end{align*}
where we have used the orthonormality condition (\ref{eq:orthoYlm}) and the fact that $Y_0^0=\frac{1}{\sqrt{4\pi}}$.

Likewise, since $Y_1^0(\theta)=\sqrt{\frac{3}{4\pi}} \cos(\theta)$, the Stark potential (\ref{eq:WStark}) can be written in spherical coordinates as
$$
W_{\rm Stark}(r,\theta)=-\sqrt{\frac{4\pi}{3}}r Y_1^0(\theta)=-\sqrt{\frac{4\pi}{3}}r Y_1^0(\theta)^\ast,
$$
and the potential energy due to the external electric field is then equal to
$$
\beta \int_{\R^3} \rho_hW_{\rm Stark} = -\frac{1}{\sqrt 3}\beta\dps\sum_{i,j=1}^{N_h}[R_1]_{ij} [M_{1}]_{ij}=  -\frac{1}{\sqrt 3}\beta \tr(M_1R_1).
$$
Let $\mu$ be a radial, continuous function from $\R^3$ to $\R$ vanishing at infinity and such that $\int_{\R^3}\mu=1$.  The Coulomb interaction energy can be rewritten as follows:
\begin{equation}
\frac 12 D(\rho_h,\rho_h) =  \frac 12 D\left(\rho_h-\left( \int_{\R^3} \rho_h \right) \mu,\rho_h-\left( \int_{\R^3} \rho_h \right) \mu\right) +  N  D(\mu,\rho_h) - \frac {N^2}2 D(\mu,\mu) . \label{eq:decD}
\end{equation}
The reason why we introduce the charge distribution $\mu$ is to make neutral the charge distributions $\rho_h-\left( \int_{\R^3} \rho_h \right) \mu$ in the first term of the right-hand side of (\ref{eq:decD}), in such a way that the physical solution $Q_{0,R_0}$ to the equation (\ref{eq:decD2}) below for $l=0$ is in $H^1_0(0,+\infty)$.

Introducing the real symmetric matrix $V_\mu \in \R^{N_h \times N_h}$ with entries
\begin{equation}\label{def:V_mu}
[V_\mu]_{ij}=\int_0^{+\infty}  [V^{\rm H}(\mu)](r\be)  \cX_i(r)\cX_j(r) \, dr,
\end{equation}
where $\be$ is any unit vector of $\R^3$ (the value of $V^{\rm H}(\mu)(r\be)$ is independent of $\be$ since $V^{\rm H}(\mu)$ is radial) the sum of the last two terms of the right-hand side of (\ref{eq:decD}) can be rewritten as 
$$
N D(\mu,\rho_h) -\frac {N^2}2 D(\mu,\mu) = N \tr(V_\mu R_0) - \frac {N^2}2 D(\mu,\mu).
$$
Denoting by 
$$
\widetilde V^{\rm H}(\rho_h)=V^{\rm H}\left( \rho_h-\left( \int_{\R^3} \rho_h \right)\mu\right),
$$
we have by symmetry $\widetilde V^{\rm H}(\rho_h) \in \cH^0$ and
$$
[\widetilde V^{\rm H}(\rho_h)](r,\theta)= \sum_{l=0}^{2m_h} \widetilde V_l(\rho^h_l)(r) Y_l^0(\theta)=\sum_{l=0}^{2m_h} \frac{Q_{l,R_l}(r)}{r} Y_l^0(\theta),
$$
where $Q_{l,R_l}$ is the unique solution in $H^1_0(0,+\infty)$ to the differential equation
 \begin{equation}\label{eq:Poisson_l}
-\frac{d^2Q_{l,R_l}}{dr^2}(r)+\frac{l(l+1)}{r^2}Q_{l,R_l}(r)=4\pi r\left( \left(\frac{1}{\sqrt{4\pi}}\sum_{i,j=1}^{N_h} [R_l]_{ij} \frac{\cX_i(r)\cX_j(r)}{r^2} \right)-N \mu(r)\delta_{l0} \right).
\end{equation}
Note that the mappings $R_l \mapsto Q_{l,R_l}$ are linear. We therefore obtain
\begin{align}
\frac 12 D(\rho_h,\rho_h) = & \frac 12 \sum_{l=0}^{2 m_h}\frac{1}{4\pi}\left( \int_0^{+\infty} \left( \left( \frac{dQ_{l,R_l}}{dr} (r) \right)^2 + \frac{l(l+1)}{r^2} Q_{l,R_l} (r)^2  \right) \, dr \right)  \nonumber \\ &  
 +N \tr(V_\mu R_0) - \frac {N^2}2 D(\mu,\mu).
 \label{eq:decD2}
\end{align}
Finally, the exchange-correlation energy is
\begin{equation}\label{eq:Excrhoh}
E_{\rm xc}(\rho_h)=2\pi \int_0^{+\infty} \int_0^{\pi} \epsilon_{\rm xc} \left( \frac{1}{\sqrt{4\pi}} \sum_{l=0}^{2m_h} \sum_{i,j=1}^{N_h} [R_l]_{ij} \frac{\cX_i(r)}{r}\frac{\cX_j(r)}{r} Y_l^0(\theta) \right) r^2 \sin\theta \, dr \; d\theta.
\end{equation}

\subsubsection{Approximation of the Hartree term}

Except for very specific basis functions (such as Gaussian atomic orbitals), it is not possible to evaluate exactly the first contribution to the Coulomb energy~(\ref{eq:decD2}). It is therefore necessary to approximate it. For this purpose, we use a variational approximation of (\ref{eq:Poisson_l})-(\ref{eq:decD2}) in an auxiliary basis set $\left\{\zeta_p\right\}_{1 \le p \le N_{h,\rm a}} \in (H^1_0(0,+\infty))^{N_{h,\rm a}}$, which amounts to replacing $\frac 12 D(\rho_h,\rho_h)$ by its lower bound
\begin{align}
\frac 12 D_h(\rho_h,\rho_h) = & \frac{1}{8\pi} \left( \int_0^{+\infty}  \left( \left( \frac{dQ_{l,R_l}^h}{dr} (r) \right)^2 + \frac{l(l+1)}{r^2} Q_{l,R_l}^h (r)^2  \right) \, dr   \right)  \nonumber \\ &  
 +N \tr(V_\mu R_0) - \frac {N^2}2 D(\mu,\mu),
 \label{eq:decD3}
\end{align}
where $Q_{l,R_l}^h$ is the unique solution in $\zeta^h=\mbox{span}(\zeta_1,\cdots,\zeta_{N_{h,\rm a}})$ to the problem
\begin{align*}
\forall v_h \in \zeta^h, \quad & \int_0^{+\infty} \left( \frac{dQ_{l,R_l}^h}{dr} (r)  \frac{dv_h}{dr}(r) + \frac{l(l+1)}{r^2} Q_{l,R_l}^h(r)  v_h (r)  \right) \, dr  \\
&=  4\pi\int_0^{+\infty} r\left(  \left(\frac{1}{\sqrt{4\pi}}\sum_{i,j=1}^{N_h} [R_l]_{ij} \frac{\cX_i(r)\cX_j(r)}{r^2} \right) -N \mu(r) \delta_{l0} \right)v_h(r)dr,
\end{align*}
which is nothing but the variational approximation of (\ref{eq:Poisson_l}) in the finite dimensional space $\zeta^h$. Expanding the functions $Q_{l,R_l}^h$ in the basis set $\left\{\zeta_k\right\}_{1 \le k \le N_{h,\rm a}}$ as
$$
Q_{l,R_l}^h(r) = \sum_{p=1}^{N_{h,\rm a}} Q_{p,l} \zeta_p(r),
$$
and collecting the coefficients $Q_{p,l}$, $1 \le k \le N_{h,\rm a}$ in a vector $Q_l \in \R^{N_{h,\rm a}}$, we obtain that the vector $Q_l$ is solution to the linear system 
\begin{equation}\label{eq:Poisson_l_2}
\left(A^{\rm a}+l(l+1)M_{-2}^{\rm a}\right)Q_l =4\pi\left(F : R_l -  N \delta_{l0} G\right),
\end{equation}
where the $N_{h,\rm a} \times N_{h,\rm a}$ real symmetric matrices $A^{\rm a}$ and $M_{-2}^{\rm a}$ are defined by
\begin{equation}\label{eq:matrices2}
A_{pq}^{\rm a}=\int_0^{+\infty} \zeta_p'\zeta_q',\quad [M_{-2}^{\rm a}]_{pq}=\int_0^{+\infty} \frac{\zeta_p(r)\zeta_q(r)}{r^2} \, dr,
\end{equation}
where $F\in\R^{N_{h,\rm a}\times N_h\times N_h}$  is the three-index tensor with entries
\begin{eqnarray}\label{tensorF}
 F_{pij}=\frac{1}{\sqrt{4\pi}} \int_0^{+\infty} \frac{\cX_i(r)\cX_j(r)\zeta_p(r)}{r} \, dr, 
\end{eqnarray}
and where $G\in\R^{N_{h,\rm a}}$ is the vector with entries
\begin{equation}\label{vectorG}
 G_p=\int_0^{+\infty} r \mu(r) \zeta_p(r) \, dr.
\end{equation}
Note that since $N=\tr(M_0R_0)$, the mappings $R_l \mapsto Q_l$ are in fact linear. We finally get  
\begin{equation}
\frac 12 D_h(\rho_h,\rho_h) =  \frac{1}{8\pi} \sum_{l=0}^{2m_h} Q_l^T ( A^{\rm a}+l(l+1)M_{-2}^{\rm a}) Q_l  
 +N \tr(V_\mu R_0) - \frac {N^2}2 D(\mu,\mu),
 \label{eq:decD4}
\end{equation}
where $Q_l$ is the solution to (\ref{eq:Poisson_l_2}).

\subsubsection{Final form of the discretized problem and Euler-Lagrange equations}

We therefore end up with the following approximation of problem (\ref{eq:min_pert}): 
\begin{align}\label{eq:min_pert_approx_2}
\widetilde\cI_{z,N,h}^{\rm rHF/LDA}(\beta W) :=& \inf\bigg\{ \cE_{z,N,\beta}^{\rm rHF/LDA}((U^{m,k}),(n_{m,k})), \; (U^{m,k})_{\underset{1\leq k\leq (m_h-|m|+1) N_h}{-m_h\leq m\leq m_h}} \in \cU_{h}, \nonumber \\ & \qquad\qquad \; (n_{m,k})_{\underset{1\leq k\leq (m_h-|m|+1) N_h}{-m_h\leq m\leq m_h}} \in \cN_{N,h} \bigg\}.
\end{align}
where
\begin{align*}
\cE_{z,N,\beta}^{\rm rHF/LDA}((U^{m,k}),(n_{m,k})) :=& \frac 12\dps\sum_{\underset{1\leq k\leq (m_h-|m|+1)N_h}{-m_h\leq m\leq m_h}} n_{m,k}
                   \dps  \left( \tr\left( [U^{m,k}]^T A U^{m,k} \right) + \tr\left( D_m [U^{m,k}]^T M_{-2} U^{m,k} \right)   \right) \\
                                      & -z \tr(M_{-1}R_0)   + \frac{1}{8\pi} \sum_{l=0}^{2m_h} Q_l^T ( A^{\rm a}+l(l+1)M_{-2}^{\rm a}) Q_l   
 + N \tr(V_\mu R_0)\\ 
& - \frac {N^2}2 D(\mu,\mu) + E_{\rm xc}(\rho_h) - \frac{\beta}{\sqrt 3} \tr(M_1R_1) ,
\end{align*}
where for each $l$, the matrix $R_l$ and the vector $Q_l$ are respectively defined by (\ref{eq:defRl}) and (\ref{eq:Poisson_l_2}), and where the last but one term in the right-hand side is given by (\ref{eq:Excrhoh}).

\medskip

The gradient of $\cE_{z,N,\beta}^{\rm rHF/LDA}$ with respect to $U^{m,k}$ is 
\begin{align*}
\nabla_{U^{m,k}} \cE_{z,N,\beta}^{\rm rHF/LDA} =& 2 n_{m,k} \bigg( \frac 12 A U^{m,k} + \frac 12 M_{-2} U^{m,k} D_m - z  M_{-1} U^{m,k} + N V_\mu  U^{m,k}  \\
 & \qquad + \sum_{l=0}^{2m_h} (Q_l^T \cdot F) (U^{m,k}C^{l,m}) + \sum_{l=0}^{2m_h} V_{\rm xc}^l U^{m,k} C^{l,m} -\frac{\beta}{\sqrt 3}   M_1 U^{m,k} C^{1,m} \bigg),
\end{align*}
where for each $0 \le l \le 2m_h$, the $N_h \times N_h$ real matrix $V_{\rm xc}^l$ is defined by
\begin{equation}\label{eq:Vxcrhoh}
[V^l_{\rm xc}]_{ij} = \sqrt \pi \, \int_0^{+\infty} \int_0^\pi v_{\rm xc}  \left(\frac{1}{\sqrt{4\pi}}\sum_{i,j=1}^{N_h} [R_l]_{ij} \frac{\cX_i(r)\cX_j(r)}{r^2}\right) \cX_i(r) \cX_j(r) Y_l^0(\theta) \, \sin\theta \, dr \, d\theta,
\end{equation}
where $v_{\rm xc}(\rho):=\frac{d \epsilon_{\rm xc}}{d\rho}(\rho)$ is the exchange-correlation potential. 

\medskip

Diagonalizing simultaneously the Kohn-Sham Hamiltonian and the ground state density matrix in an orthonormal basis, we obtain that the ground state can be obtained by solving the following system of first-order optimality conditions, which is nothing but a reformulation of the discretized extended Kohn-Sham equations exploiting the cylindrical symmetry of the problem:
\begin{align} 
&\dps \frac 12 A U^{m,k} + \frac 12 M_{-2} U^{m,k} D_m - z  M_{-1} U^{m,k}  + N V_\mu  U^{m,k} + \sum_{l=0}^{2m_h} (Q_l^T \cdot F) (U^{m,k}C^{l,m}) \nonumber
\\ 
&\qquad \qquad \dps  + \sum_{l=0}^{2m_h} V^l_{\rm xc} U^{m,k} C^{l,m}-\frac{1}{\sqrt 3}  \beta  M_1 U^{m,k} C^{1,m} =  \epsilon_{m,k} M_0 U^{m,k},  \label{eq:EKSequations-1}  \\  \nonumber \\
&\dps \tr\left( [U^{m,k}]^T M_0 U^{m,k'} \right) = \delta_{kk'}, \\  \nonumber \\ 
&\dps \left(A^{\rm a}+l(l+1)M_{-2}^{\rm a}\right)Q_l =F : R_l -  \tr(M_0R_0)  \delta_{l0} G, \\  \nonumber \\
& \!\!\!\!\!\!\!\!\!\!\!\!\!\!\!\!\!\!\!\!\!\!\! \dps [V^l_{\rm xc}]_{ij} = \sqrt \pi \, \int_0^{+\infty} \int_0^\pi v_{\rm xc}  \left(\frac{1}{\sqrt{4\pi}}\sum_{i,j=1}^{N_h} [R_l]_{ij} \frac{\cX_i(r)\cX_j(r)}{r^2}\right) \cX_i(r) \cX_j(r) Y_l^0(\theta) \, \sin\theta \, dr \, d\theta, \\  \nonumber \\
&\dps  n_{m,k}=2 \mbox{ if } \epsilon_{m,k}< \epsilon_{\rm F}, \quad
 0\leq n_{m,k}\leq 2\mbox{ if } \epsilon_{m,k}= \epsilon_{\rm F}, \quad
 n_{m,k}=0 \mbox{ if } \epsilon_{m,k}> \epsilon_{\rm F}, \\  \nonumber \\
& \dps \sum_{\underset{1\leq k\leq (m_h-|m|+1) N_h}{-m_h\leq m\leq m_h}}n_{m,k}=N, \\  \nonumber \\
& \dps R_l = \sum_{\underset{1\leq k\leq (m_h-|m|+1) N_h}{-m_h\leq m\leq m_h}}n_{m,k} U^{m,k} C^{l,m} [U^{m,k}]^T,\label{eq:EKSequations-7} 
\end{align}
where the matrices $A$, $M_{-2}$, $M_{-1}$, $M_0$, $M_1$, $D_m$, $V_\mu$, $A^{\rm a}$, $M_{-2}^{\rm a}$, $C^{l,m}$, the 3-index tensor $F$ and the vector $G$ are defined by (\ref{matrices}), (\ref{eq:matDm}), (\ref{eq:matrixClm}), (\ref{def:V_mu}), (\ref{eq:matrices2}), (\ref{tensorF}), (\ref{vectorG}).

\subsubsection{$\P_4$-finite element method}\label{sec:P4}

In our calculations, we use the same approximation space to discretize the radial components of the Kohn-Sham orbitals and the radial Poisson equations (\ref{eq:Poisson_l}), so that, in our implementation of the method, $N_{h,\rm a}=N_h$ and $\cX^h=\zeta^h$. We choose a cut-off radius $L_e > 0$ large enough and discretize the interval $[0,L_e]$ using a non-uniform grid with $N_I+1$ points $0=r_1 < r_2 < \cdots < r_{N_I} < r_{N_I+1}=L_e$. The positions of the points are chosen according to the following rule:
$$
r_k=r_{k-1}+h_k, \qquad h_{N_I}=\frac{1-s}{1-s^{N_I}} L_e, \qquad h_{k-1}=s h_k,
$$
where $0 < s < 1$ is a scaling parameter leading to a progressive refinement of the mesh when one gets closer to the nucleus ($r=0)$. To achieve the desired accuracy, we use the $\P_4$-finite element method.  

\medskip

All the terms in the variational discretization of the energy and of the constraints can be computed exactly (up to finite arithmetics errors), except the exchange-correlation terms (\ref{eq:Excrhoh}) and (\ref{eq:Vxcrhoh}), which requires a numerical quadrature method. In our calculation, we use Gaussian quadrature formulas~\cite{gauss-quad} of the form
\begin{align*}
\int_0^{+\infty}\int_0^\pi f(r,\theta) \sin\theta\, dr \, d\theta&=\int_0^{+\infty}\int_{-1}^1 f(r,\arccos t_\theta)  \, dr \, dt_\theta  \\
&\simeq \sum_{k=1}^{N_I} \sum_{i=1}^{N_{{\rm g},r}} \sum_{j=1}^{N_{{\rm g},\theta}}   h_k w_{i,\rm r} w_{j,\theta}  f(r_k+h_k t_{i,\rm r}, \arccos(t_{j,\theta})), 
\end{align*}

where the $0 < t_{1,r} < \cdots < t_{N_{{\rm g},r},r} < 1$ (resp. $-1 < t_{1,\theta} < \cdots < t_{N_{{\rm g},\theta},\theta} < 1$) are Gauss points for the $r$-variable (resp. for the $t_\theta$-variable) with associated weights $w_{1,r}, \cdots , w_{N_{{\rm g},r},r}$ (resp.  $w_{1,\theta}, \cdots , w_{N_{{\rm g},\theta},\theta}$). 

\medskip

More details about the practical implementation of the method are provided in Appendix.

\subsection{Description of the algorithm}
In order to solve the self-consistent equations (\ref{eq:EKSequations-1})-(\ref{eq:EKSequations-7}), we use an iterative algorithm. For clarity, we first present this algorithm within the continuous setting. Its formulation in the discretized setting considered here is detailed below. The iterations are defined as follows: an Ansatz of the ground state density $\rho^{[n]}$ being known,
\begin{enumerate}
\item construct the Kohn-Sham operator 
$$
H^{[n]} = -\frac 12 \Delta -\frac z{|\cdot|}+V^{\rm H}(\rho^{[n]})+v_{\rm xc}(\rho^{[n]})+\beta W
$$
where $v_{\rm xc}=0$ for the rHF model and $v_{\rm xc}=v_{\rm xc}^{\rm LDA}$ for the Kohn-Sham LDA model;
\item for each $m \in \Z$, compute the negative eigenvalues of $H^{[n]}_m:=\Pi_m H^{[n]} \Pi_m$, where $\Pi_m$ is the orthogonal projector on the space $\cH^m$:
$$
H^{[n]}_m \phi_{m,k}^{[n+1]}  = \epsilon_{m,k}^{[n+1]} \phi_{m,k}^{[n+1]}, \qquad \int_{\R^3}  {\phi_{m,k}^{[n+1]}}^\ast \phi_{m,k'}^{[n+1]} = \delta_{kk'};
$$
\item construct a new density 
$$
\rho^{[n+1]}_* = \sum_{m,k} n_{m,k}^{[n+1]} |\phi_{m,k}^{[n+1]}|^2,
$$
where
$$
\left\{\begin{array}{lcl}
 n_{m,k}^{[n+1]}=2 &\mbox{ if }& \epsilon_{m,k}^{[n+1]} < \epsilon_{\rm F}^{[n+1]}, \\
 0\leq n_{m,k}^{[n+1]}\leq 2 &\mbox{ if }& \epsilon_{m,k}^{[n+1]} = \epsilon_{\rm F}^{[n+1]}, \\
 n_{m,k}^{[n+1]}=0 &\mbox{ if }& \epsilon_{m,k}^{[n+1]} > \epsilon_{\rm F}^{[n+1]},
\end{array}\right. \qquad \mbox{and} \qquad \sum_{(m,k)} n_{m,k}^{[n+1]}=N;
$$
\item update the density:
$$
\rho^{[n+1]}=t_{n} \rho^{[n+1]}_* + (1-t_{n}) \rho^{[n]},
$$
where $t_{n}\in[0,1]$ either is a fixed parameter independent of $n$ and chosen \textit{a priori}, or is optimized using the Optimal Damping Algorithm (ODA), see below;
\item if some convergence criterion is satisfied, then stop; else, replace $n$ with $n+1$ and go to step 1.
\end{enumerate}
In the non-degenerate case, that is when $\epsilon_{\rm F}^{[n+1]}$ is not an eigenvalue of the Hamiltonian $H^{[n]}$, the occupation numbers $n_{m,k}^{[n+1]}$ are equal to either $0$ (unoccupied) or $2$ (fully occupied), while in the degenerate case the occupation numbers at the Fermi level have to be determined. We distinguish two cases: if $W=0$, or more generally if $W$ is spherically symmetric, and if $\epsilon_{\rm F}^{[n+1]}$ is not an accidentally degenerate eigenvalue of $H^{[n]}$, then the occupation numbers 
at the Fermi level are all equal; otherwise, the occupation numbers are not known \textit{a priori}. In our approach we select the occupation numbers at the Fermi level which provide the lowest Kohn-Sham energy.  When the degenerate eigenspace at the Fermi level is of dimension $3$, that is when the highest energy partially occupied orbitals are perturbations of a three-fold degenerate p-orbital, the optimal occupation numbers can be found by using the golden search or bisection method~\cite[Chapter 10]{numerical-recipes} since, in this case, the search space can be parametrized by a single real-valued parameter (this is due to the fact that the sum of the three occupation numbers is fixed and that two of them are equal by cylindrical symmetry). In the general case, more generic optimization methods have to be resorted to.

\medskip

In the discretization framework we have chosen, the algorithm can be formulated as follows.

\medskip

\noindent
{\bf Initialization.} 
\begin{enumerate}
\item Choose the numerical parameters $m_h$ (cut-off in the spherical harmonics expansion), $L_e$  (size of the simulation domain for the radial components of the Kohn-Sham orbitals and the electrostatic potential), $N_I$ (size of the mesh for solving the radial equations), $N_{{\rm g},r}$ (number of Gauss points for the radial quadrature formula), $N_{{\rm g},\theta}$ (number of Gauss points for the angular quadrature formula), and $\varepsilon > 0$ (convergence threshold),
\item assemble the matrices $A=A^{\rm a}$, $M_{-2}=M_{-2}^{\rm a}$, $M_{-1}$, $M_0$, $M_1$, $C^{l,m}$, $V_\mu$ and the vector $G$. The tensor $F$ can be either computed once and for all, or the contractions $F:R_l^{[n]}$ can be computed on the fly, depending on the size of the discretization parameters and the computational means available;
\item choose an initial guess $(R_l^{[0]})_{0 \le l\le 2m_h}$ for the matrices representing the discretized ground state density at iteration $0$ (it is possible to take $R_l=0$ for all $l$ if no other better guess is known).
\end{enumerate}

\medskip

\noindent
{\bf Iterations.} The matrices $(R_l^{[n]})_{0 \le l\le 2m_h}$ at iteration $n$ being known,
\begin{enumerate}
\item construct the building blocks of the discretized analogues of the operators $H^{[n]}_m$. For this purpose,
\begin{enumerate}
\item solve, for each $l=0,\cdots,2m_h$, the linear equation
 $$
 \left(A^{\rm a}+l(l+1)M_{-2}^{\rm a}\right)Q_l^{[n]} =4\pi\left(F : R_l^{[n]} -  N \delta_{l0} G\right)
 $$
 \item assemble, for each $l=0,\cdots,2m_h$, the matrix $V_l^{{\rm xc},[n]}$ by means of use Gauss quadrature rules
 $$
  [V^{l,[n]}_{\rm xc}]_{ij} = \sqrt \pi \sum_{k=1}^{N_I} \sum_{p=1}^{N_{{\rm g},r}} \sum_{q=1}^{N_{{\rm g},\theta}}   h_k w_{p,\rm r} w_{q,\theta}  f_{ij}^l(r_k+h_k t_{p,\rm r}, t_{q,\theta}),
 $$
 where
 $$
 \!\!\!\!\!\!\!\!f_{ij}^l(r,t_\theta) = v_{\rm xc}  \left(\frac{1}{\sqrt{4\pi}}\sum_{l=0}^{m_h}\sum_{i,j=1}^{N_h} [R_l]_{i,j} \frac{\cX_i(r)\cX_j(r)}{r^2}Y_l^0(\arccos t_\theta)\right) \cX_i(r) \cX_j(r) Y_l^0(\arccos t_\theta);
 $$
 \end{enumerate} 
\item solve, for each ${0} \le m \le m_h$, the generalized eigenvalue problem
\begin{align} \label{eq:EKSequationsn} 
&\dps  \!\!\!\!\!\!\!\!\!\!\!\!\!\!\!\!\!\!\!\!\!\!  \frac 12 A U^{m,k,[n+1]}  + \frac 12 M_{-2} U^{m,k,[n+1]} D_m - z  M_{-1} U^{m,k,[n+1]}  + N V_\mu  U^{m,k,[n+1]} + \sum_{l=0}^{2m_h} (Q_l^{[n]T} \cdot F) (U^{m,k,[n+1]}C^{l,m}) \nonumber
\\ 
&\qquad \qquad \dps  + \sum_{l=0}^{2m_h} V_{{\rm xc}}^{l,[n]} U^{m,k,[n+1]} C^{l,m}-\frac{\beta}{\sqrt 3}    M_1 U^{m,k,[n+1]} C^{1,m} =  \epsilon_{m,k}^{[n+1]} M_0 U^{m,k,[n+1]},   \\  \nonumber \\
&\dps \tr\left( [U^{m,k,[n+1]}]^T M_0 U^{m,k',[n+1]} \right) = \delta_{kk'},  \label{eq:constraintsn} 
\end{align}
 \item build the matrices $R_{l,*}^{[n+1]}$ using the \textit{Aufbau} principle and, if necessary, optimizing the occupation numbers $n_{m,k}^{[n+1]}$, by selecting the occupation numbers at the Fermi level leading to the lowest Kohn-Sham energy\footnote{In practice, this optimization problem is low-dimensional. Indeed, the degeneracy of the Fermi level is typically 3 (perturbation of p-orbitals) or 5 (perturbation of d-orbitals) for most atoms of the first four rows of the periodic table, and some of the occupation numbers are known to be equal for symmetric reasons.}:
 $$
 R_{l,*}^{[n+1]} =  \sum_{\underset{1\leq k\leq (m_h-|m|+1) N_h}{-m_h\leq m\leq m_h}}n_{m,k}^{[n+1]} U^{m,k,[n+1]} C^{l,m} [U^{m,k,[n+1]}]^T, 
 $$
 where
 $$
\left\{\begin{array}{lcl}
 n_{m,k}^{[n+1]}=2 &\mbox{ if }& \epsilon_{m,k}^{[n+1]} < \epsilon_{\rm F}^{[n+1]}, \\
 0\leq n_{m,k}^{[n+1]}\leq 2 &\mbox{ if }& \epsilon_{m,k}^{[n+1]} = \epsilon_{\rm F}^{[n+1]}, \\
 n_{m,k}^{[n+1]}=0 &\mbox{ if }& \epsilon_{m,k}^{[n+1]} > \epsilon_{\rm F}^{[n+1]},
\end{array}\right. \qquad \mbox{and} \qquad \sum_{(m,k)} n_{m,k}^{[n+1]}=N;
$$
 \item update the density: 
 $$
 \forall 0 \le l \le 2m_h, \quad R_l^{[n+1]}=t_n R_{l,*}^{[n+1]}+(1-t_n)R_l^{[n]},
 $$
 where $t_n\in[0,1]$ either is a fixed parameter independent of $n$ and chosen 
       \textit{a priori}, or is optimized using the ODA, see below;
 \item if (for instance) $\max_{0 \le l \le 2m_h} \| R_l^{[n+1]} - R_l^{[n]} \| \le \epsilon$ or $|E^{[n+1]}-E^{[n]}| \leq \varepsilon$ then stop; else go to step one.  
\end{enumerate}

Note that the generalized eigenvalue problem (\ref{eq:EKSequationsn})-(\ref{eq:constraintsn}) can be rewritten as a standard generalized eigenvalue problem of the form
\begin{equation}\label{eq:GEP}
\H^m \V_k = \epsilon_{m,k}^{[n+1]} \M \V_k, \qquad \V_k^T \M \V_{k'} = \delta_{kk'},
\end{equation}
where the unknowns are vectors (and not matrices) by introducing the column vectors $\V_k \in \R^{(m_h+1-|m|)N_h}$  and the block matrices
$$
\H^m\in \R^{(m_h+1-|m|)N_h \times (m_h+1-|m|)N_h} \qquad \mbox{and} \qquad \M\in \R^{(m_h+1-|m|)N_h \times (m_h+1-|m|)N_h}
$$
defined as
$$
\V_k = \left( \begin{array}{c} U^{m,k,[n+1]}_{\cdot,|m|}   \\    \hline \cdot \\ \cdot \\ \cdot \\ \hline U^{m,k,[n+1]}_{\cdot,m_h} \end{array} \right), \qquad \H^m = \left( \begin{array}{c|c|c|c|c}  \H^m_{|m|,|m|} & \H^m_{|m|,|m|+1}   & \cdots  & \H^m_{|m|,m_h-1,}  &  \H^m_{|m|,m_h} \\  \hline   \H^m_{|m|+1,|m|} & \H^m_{|m|+1,|m|+1}   & \cdots  & \H^m_{|m|+1,m_h-1,}  & \H^m_{|m|+1,m_h}   \\ \hline  \cdots & \cdots & \cdots & \cdots & \cdots  \\ \hline
 \H^m_{m_h-1,|m|} & \H^m_{m_h-1,|m|+1}   & \cdots  & \H^m_{m_h-1,m_h-1,}  &  \H^m_{m_h-1,m_h} \\  \hline   \H^m_{m_h,|m|} & \H^m_{m_h,|m|+1}   & \cdots  & \H^m_{m_h,m_h-1,}  & \H^m_{m_h,m_h}  
 \end{array} \right),
$$
and
$$
\M = \mbox{block diag}(M_0,\cdots,M_0),
$$
where each of the $(m_h-|m|+1)$ block $\H^m_{l,l'}$ is of size $N_h \times N_h$ with
$$
\forall |m| \le l \le m_h, \quad \H^m_{l,l} =   \frac 12 A + \frac{l(l+1)}2 M_{-2} -z M_{-1}+NV_\mu  + \sum_{l''=0}^{2m_h} C^{l,m}_{l,l''} \left([Q_{l''}^{[n]}]^T\cdot F+V_{\rm xc}^{l'',[n]}\right)
$$
$$
\forall |m| \le l \neq l' \le m_h, \quad \H^m_{l,l'} =    \sum_{l''=0}^{2m_h} C^{l,m}_{l',l''} \left([Q_{l''}^{[n]}]^T\cdot F+V_{\rm xc}^{l'',[n]}\right) -\frac{\beta}{\sqrt 3} C^{1,m} M_1 \delta_{|l-l'|,1}.
$$

If $\beta=0$ and if the density $\rho_h^{[n]}$ is radial, then $R_l^{[n]}=0$ for all $l \in \N^\ast$, and the matrix $\H^m$ is block diagonal. The generalized eigenvalue problem (\ref{eq:GEP}) can then be decoupled in $(m_h-|m|+1)$ independent generalized eigenvalue problems of size $N_h$. This comes from the fact that the problem being spherically symmetric, the Kohn-Sham Hamiltonian is block diagonal in the two decompositions 
$$
L^2(\R^3) = \bigoplus_{l \in \N} \cH_l \qquad \mbox{and} \qquad L^2(\R^3) = \bigoplus_{m \in \Z} \cH^m.
$$

\medskip
Let us conclude this section with some remarks on the Optimal Damping Algorithm (ODA)~\cite{CB2,CB1}, used to find an optimal step-length $t_n$ to mix the matrices $R_{l,*}^{[n+1]}$ and $R_l^{[n]}$ in Step 4 of the iterative algorithm. This 
step-length is obtained by minimizing on the range $t\in[0,1]$ the one-dimensional function 
$$
t\mapsto \widetilde E_{z,N}^{\rm rHF/LDA} \left((1-t)\gamma^{[n+1]}_*+t\gamma^{[n]},\beta W\right),
$$
where $\gamma^{[n]}$ is the current approximation of the ground state density matrix at iteration $n$ and
$$
\gamma^{[n+1]}_*= \dps\sum_{\underset{1\leq k\leq (m_h-|m|+1)N_h}{-m_h\leq m\leq m_h}}n_{m,k}^{[n+1]} |\Phi_{m,k,h}^{[n+1]}\rangle\langle\Phi_{m,k,h}^{[n+1]}|,
$$
with
$$
\Phi_{m,k,h}^{[n+1]}(r,\theta,\varphi)=\sum_{l=|m|}^{m_h} \sum_{i=1}^{N_h} U_{i,l}^{m,k,[n+1]}   \frac{\cX_i(r)}{r} Y_{l}^{m}(\theta,\varphi),
$$
A key observation is that this optimization problem can be solved without storing density matrices, but only the two sets of matrices $R^{[n]}:=(R_l^{[n]})_{0 \le l \le 2m_h}$ and $R_*^{[n+1]}:=(R_{l,*}^{[n+1]})_{0 \le l \le 2m_h}$, and the scalars
$$
E_{{\rm kin}}^{[n]} := \tr \left( -\frac 12 \Delta \gamma^{[n]}\right)
$$
and
\begin{align*} 
E_{{\rm kin},*}^{[n+1]} &:= \tr \left( -\frac 12 \Delta \gamma^{[n+1]}_*\right) \\ &= \frac 12\dps\sum_{\underset{1\leq k\leq (m_h-|m|+1)\times N_h}{-m_h\leq m\leq m_h}} n_{m,k}^{[n+1]}
                   \dps  \left( \tr\left( [U^{m,k,[n+1]}]^T A U^{m,k,[n+1]} \right) \right.\\& \left.\qquad\qquad\qquad\qquad\qquad+ \tr\left( D_m [U^{m,k,[n+1]}]^T M_{-2} U^{m,k,[n+1]} \right)   \right).
\end{align*}
Indeed, we have for all $t \in [0,1]$, 
$$
\widetilde E_{z,N}^{\rm rHF/LDA} \left((1-t)\gamma^{[n+1]}_*+t\gamma^{[n]},\beta W\right) =   (1-t) E_{{\rm kin},*}^{[n+1]} +t  E_{{\rm kin}}^{[n]}   + {\cal F}^{\rm rHF/LDA} \left( (1-t)R_{*}^{[n+1]} +t R^{[n]},\beta W \right),
$$
where the functional ${\cal F}^{\rm rHF/LDA}$ collects all the terms of the Kohn-Sham functional depending on the density only.
When $E_{\rm xc}=0$ (rHF model), the function 
$$t \mapsto \widetilde E_{z,N}^{\rm rHF/LDA} \left((1-t)\gamma^{[n+1]}_*+t\gamma^{[n]},\beta W\right)$$
 is a convex polynomial of degree two, and its minimizer on $[0,1]$ can therefore be easily computed explicitly. In the LDA case, the minimum on $[0,1]$ of the above function of $t$ can be obtained using any line search method. We use here the golden search method
. Once the minimizer $t_n$ is found, the quantity $E^{[n]}_{\rm kin}$ is updated using the relation
$$
E^{[n+1]}_{\rm kin} =  (1-t_n) E_{{\rm kin},*}^{[n+1]} +t_n  E_{{\rm kin}}^n.
$$

\section{Numerical results}\label{sec:results}
As previously mentioned, we use in our code, written in Fortran 95 language, the same basis to discretize the radial components of the Kohn-Sham orbitals and of the Hartree potential, that is $(\cX_i)_{1\leq i\leq N_h}=(\zeta_i)_{1\leq i\leq N_h}$, and the $\P_4$ finite elements method to construct the discretization basis. 

\medskip

In order to test our methodology on LDA-type models, we have chosen to work with the X$\alpha$ model~\cite{Slater}, which has a simple analytic expression: 
$$
E_{\rm xc}(\rho)=-\frac 34\left(\frac 3\pi\right)^{\frac 13}\int_{\R^3} \rho^{\frac 43} \quad \mbox{and} \quad v_{\rm xc}(\rho)=-\left(\frac 3\pi \right)^{\frac 13} \rho^{\frac 13}. 
$$
The exchange-correlation contributions must be computed by numerical quadratures. We use here the Gauss quadrature method with $N_{{\rm g},r}=15$ and $N_{{\rm g},\theta}=30$ (see Section~\ref{sec:P4}). 

\medskip

\begin{figure}[h]
\hspace{-5mm}
\begin{tabular}{cc}
\includegraphics[width=5.2cm,angle=270]{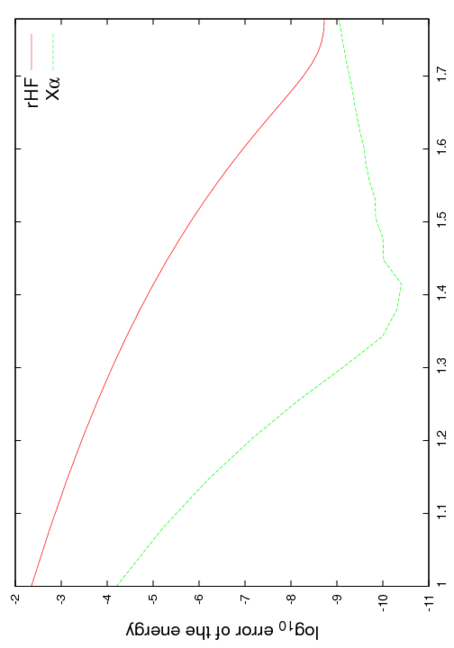} &
\includegraphics[width=5.2cm,angle=270]{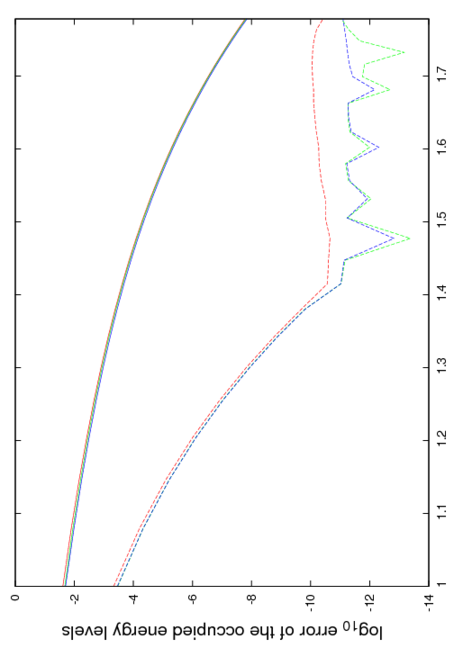}
\end{tabular}
\caption{\scriptsize Log-log plot of the error on the total energy (left) and the three occupied energy levels (right) of the carbon atom for the rHF (solid lines) and ${\rm X}\alpha$ (dashed lines) models as a function of the cut-off radius $L_e$  for a fixed mesh size $N_I=50$ (the reference calculation corresponds to $L_e=100$ and $N_I=100$).}
\label{fig:R-errors}
\end{figure}

\begin{figure}[h]
\hspace{-5mm}
\begin{tabular}{cc}
\includegraphics[width=5.2cm,angle=270]{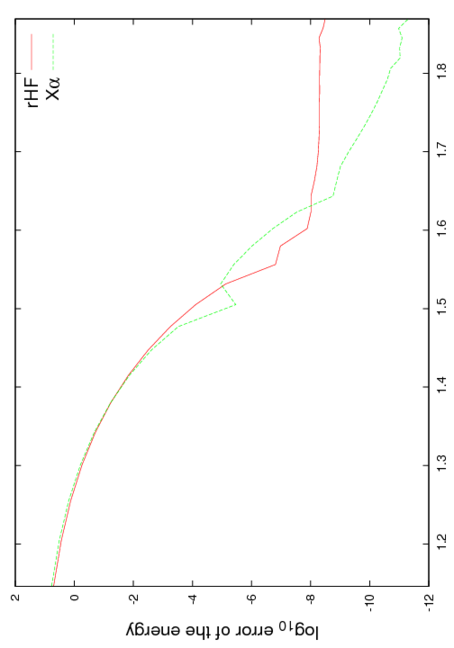} &
\includegraphics[width=5.2cm,angle=270]{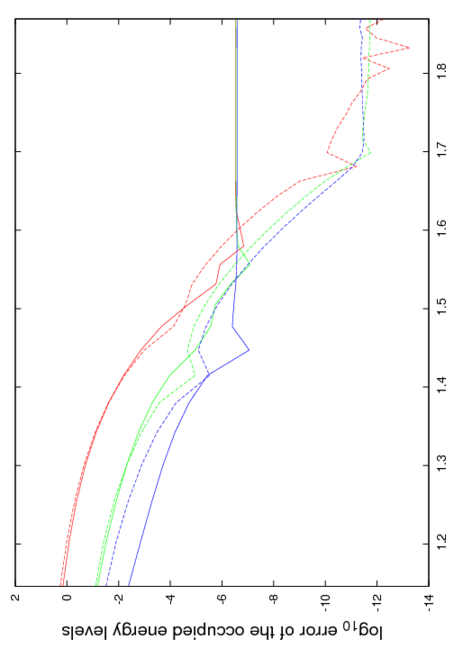}
\end{tabular}
\caption{\scriptsize Log-log plot of the error on the total energy (left) and the three occupied energy levels (right) of the carbon atom for the rHF (solid lines) and ${\rm X}\alpha$ (dashed lines) models as a function of the mesh size $N_I$, for a fixed cut-off radius $L_e=50$ (the reference calculation corresponds to $L_e=100$ and $N_I=100$).}
\label{fig:Ng-errors}
\end{figure}

We start this section by studying the convergence rate of the ground state energy and of the occupied energy levels of the carbon atom ($z=6$) as functions of the cut-off radius $L_e$ and the mesh size $N_I$ (see Section~\ref{sec:P4}). The errors on the total energy and on the occupied energy levels for the rHF and ${\rm X}\alpha$ models are plotted in Fig.~\ref{fig:R-errors} (for $L_e=50$ and different values of $N_I$) and Fig.~\ref{fig:Ng-errors} (for $N_I=50$ and different values of $L_e$), the reference calculation corresponding to $L_e=100$ and $N_I=100$. We can see that the choice $L_e=50$ and $N_I=50$ provide accuracies of about $1 \;\mu$Ha (recall that chemical accuracy corresponds to 1 mHa).

\medskip

\medskip

\subsection{Electronic structures of isolated atoms}

We report here calculations on all the atoms of the first four rows of the periodic table obtained with the rHF (Section~\ref{sec:numrHF}) and X$\alpha$  (Section~\ref{sec:numXa}) models respectively.

\subsubsection{Occupied energy levels in the rHF model}
\label{sec:numrHF}

The negative eigenvalues of $H_{\rho^0}^{\rm rHF}$ for all $1 \le z \le 54$ (first four rows of the periodic table) are listed in the tables below.
The results for $1\le z\le 20$, $27\le z\le 39$, $43\le z\le 45$ and $48\le z\le 54$  correspond to $N_I$ increasing from $35$ to $75$ as $z$ increases and  $L_e$ increasing from $30$ to $100$ as $|\epsilon_{z,z,\rm F}^{0,\rm rHF}|$ decreases, which were sufficient to obtain an accuracy of $1 \;\mu$Ha.
The remaining atoms are more difficult to deal with because the Fermi level seems to be an accidentally degenerate eigenvalue of $H_{\rho^0}^{\rm rHF}$ associated with
\begin{itemize}
\item the 4p and 3d shells for $z=21$ and $z=22$;
\item the 5s and 3d shells for $23 \le z \le 26$, with a Fermi level very close (or possibly equal) to zero;
\item the 5p and 4d shells for $z=40$, with a Fermi level very close (or possibly equal) to zero;
\item the 6s and 4d shells for $z=41$ and $z=42$, with a Fermi level very close (or possibly equal) to zero;
\item the 5s and 4d shells for $z=46$ and $z=47$.
\end{itemize}
Since the radial component of the highest occupied orbital typically vanishes as $e^{-\sqrt{2|\epsilon_{z,z,\rm F}^{0,\rm rHF}|}r}$ if $\epsilon_{z,N,\rm F}^{0,\rm rHF} < 0$ and algebraically if $\epsilon_{z,z,\rm F}^{0,\rm rHF} = 0$, a very large value of $L_e$ is needed  for the atoms for which the Fermi level is very close or possibly equal to zero.  For that case, we use a non-uniform grid with $N_I'=80$ and $L_e'=100$ as explained in Section~\ref{sec:P4} and glue to it a uniform one with $10$ points and length $L_e-L_e'$ varying from $70$ to $700$. Lastly, we add to the basis a function with an unbounded support, equal to $L_e/r$ on $[L_e,+\infty)$ (see Appendix for details). This was sufficient to obtain an accuracy of $10 \;\mu$Ha.

\medskip
 
When the accidental degeneracy involves an $s$-shell and since the density is radial, the problem of finding the occupation numbers at the Fermi level reduces to finding a single parameter $t_0 \in [0,1]$, which encodes the amount of electrons on the upper $s$-shell. In other words, one can write 
$$
\rho^{0,\rm rHF}_{z,z}=\rho_{\rm f}+t_0\rho_{\rm s}+(1-t_0)\rho_{\rm d},
$$
where $\rho_{\rm f}$ is the density corresponding to the fully occupied shells, and where $\rho_{\rm s}$ and $\rho_{\rm d}$ are densities corresponding to the accidentally degenerate s and d shells. Using the same trick for accidentally degenerate p and d shells, we manage to obtain a self-consistent solution to the rHF equations, which is necessarily a ground state since the rHF model is convex in the density matrix.

\medskip

In the following tables, we report the rHF occupied energy levels (in Ha) of all the atoms of the first four rows of the periodic table. In some cases, the Fermi level seems to be an accidentally degenerate eigenvalue: 
\begin{itemize}
\item the 4p and 3d orbitals have the same energy for $z=21, 22$;
\item the 5s and 3d orbitals have the same energy for $23 \le z \le 26$;
\item the 5p and 4d orbitals have the same energy for $z=40$;
\item the 6s and 4d orbitals have the same energy for $z=41,42$;
\item the 5s and 4d orbitals have the same energy for $z=46,47$.
\end{itemize}
In all these cases, the occupation number $0 \le n \le 2$ of the partially occupied d orbitals is also given.

\medskip

\noindent
\textbf{Hydrogen and helium:}
\begin{center}
\tiny
 \begin{tabular}{|c|c|c|}
\hline
 Atom& z& 1s \\
\hline\hline
 H &1   &     -0.046222 \\ 
 \hline
 He&2   &     -0.184889 \\ 
\hline 
\end{tabular}
\end{center}

\noindent
\textbf{First row:}
\begin{center}
\tiny
\begin{tabular}{|c|c|c|c|c|}
\hline
  Atom& z&1s&2s&2p\\
\hline\hline
 Li  &3   &     -1.202701 &       -0.013221 &-\\ 
 \hline
 Be  &4   &     -2.902437 &       -0.043722 &-\\ 
 \hline
 B   &5   &     -5.407212 &       -0.164961 &       -0.002389 \\ 
 \hline
 C   &6   &     -8.555732 &       -0.265682 &       -0.012046 \\ 
 \hline
 N   &7   &    -12.390177 &       -0.384699 &       -0.027312 \\ 
 \hline
 O   &8   &    -16.912538 &       -0.522883 &       -0.047280 \\ 
 \hline
 F   &9   &    -22.123525 &       -0.680479 &       -0.071663 \\ 
 \hline
Ne &10&    -28.023481 &       -0.857597 &       -0.100342 \\ 
 \hline
\end{tabular}
\end{center}

\noindent
\textbf{Second row:}
\begin{center}
\tiny
 \begin{tabular}{|c|c|c|c|c|c|c|}
  \hline
   Atom&z& 1s&2s&2p&3s&3p\\
 \hline\hline
Na &11&    -35.065314 &       -1.453872 &       -0.514340 &       -0.012474 &-\\ 
 \hline
Mg &12&    -42.963178 &       -2.169348 &       -1.037891 &       -0.034036 &-\\ 
 \hline
Al &13&    -51.833760 &       -3.118983 &       -1.789953 &       -0.135543 &       -0.002486 \\ 
 \hline
Si &14&    -61.532179 &       -4.160128 &       -2.629056 &       -0.208803 &       -0.010768 \\ 
 \hline
P  &15&    -72.083951 &       -5.319528 &       -3.582422 &       -0.284199 &       -0.023431 \\ 
 \hline
S  &16&    -83.489746 &       -6.598489 &       -4.651551 &       -0.363585 &       -0.039746 \\ 
 \hline
Cl &17&    -95.749535 &       -7.997404 &       -5.836930 &       -0.447628 &       -0.059401 \\ 
 \hline
Ar &18&   -108.863191 &       -9.516434 &       -7.138772 &       -0.536669 &       -0.082233 \\ 
 \hline
 \end{tabular}
\end{center}

\noindent
\textbf{Third row:}
\begin{center}
\tiny\setlength\tabcolsep{2pt}
 \begin{tabular}{|c|c|c|c|c|c|c|c|}
 \hline
  Atom&{z} &1s&2s&2p &3s&3p&4s\\
 \hline\hline
K  &19&   -123.093717 &      -11.413369 &       -8.815789 &       -0.866180 &       -0.326113 &       -0.009500 \\ 
 \hline
Ca &20&   -138.233855 &      -13.478564 &      -10.658837 &       -1.225936 &       -0.596554 &       -0.024275 \\ 
 \hline
 \end{tabular}
\end{center} 

\begin{center}
\tiny\setlength\tabcolsep{2pt}
 \begin{tabular}{|c|c|c|c|c|c|c|c|c|c|c|}
  \hline
   Atom&{z}&1s&2s&2p&3s&3p&4s&4p&3d& $n$ (3d) \\
 \hline\hline
Sc &21&    -154.35864 &       -15.78538 &       -12.74151 &        -1.69002 &        -0.96964 &        -0.08646 &        -0.00262 &        -0.00262 &           0.0056  \\
 \hline
Ti &22&    -171.13186 &       -17.95490 &       -14.69008 &        -1.91684 &        -1.11529 &        -0.08224 &        -0.00056 &        -0.00056 &           0.3076  \\
 \hline
 \end{tabular}
\end{center} 

\begin{center}
\tiny\setlength\tabcolsep{2pt}
 \begin{tabular}{|c|c|c|c|c|c|c|c|c|c|c|}
  \hline
   Atom&{z}&1s&2s&2p&3s&3p&4s&5s&3d& $n$ (3d) \\
 \hline\hline
V  &23&    -188.77080 &       -20.24077 &       -16.75392 &        -2.15109 &        -1.26708 &        -0.07796 &        -0.00044 &        -0.00044 &           0.5662  \\
 \hline
Cr &24&    -207.27457 &       -22.64280 &       -18.93275 &        -2.39225 &        -1.42402 &        -0.07027 &        -0.00021 &        -0.00021 &           0.7794  \\
 \hline
Mn &25&    -226.64207 &       -25.15938 &       -21.22503 &        -2.63884 &        -1.58451 &        -0.06385 &        -0.00008 &        -0.00008 &           0.9886  \\
 \hline
Fe &26&    -246.87446 &       -27.79250 &       -23.63263 &        -2.89238 &        -1.74975 &        -0.05831 &        -0.00001 &        -0.00001 &           1.1957  \\
 \hline
 \end{tabular}
\end{center}

\begin{center}
\tiny\setlength\tabcolsep{2pt}
 \begin{tabular}{|c|c|c|c|c|c|c|c|c|}
 \hline
  Atom&{z} &1s&2s&2p&3s&3p&4s&3d\\
 \hline
 \hline
Co &27&   -267.97363 &      -30.54468 &      -26.15798 &       -3.15502 &       -1.92172 &       -0.05438 &       -0.00121 \\ 
 \hline
Ni &28&   -289.94364 &      -33.42047 &      -28.80557 &       -3.43107 &       -2.10456 &       -0.05459 &       -0.00722 \\ 
 \hline
Cu &29&   -312.78019 &      -36.41574 &      -31.57124 &       -3.71624 &       -2.29392 &       -0.05539 &       -0.01370 \\ 
 \hline
Zn &30&   -336.48301 &      -39.53045 &      -34.45491 &       -4.01038 &       -2.48957 &       -0.05646 &       -0.02026 \\ 
 \hline
 \end{tabular}
\end{center}  

\begin{center}
\tiny\setlength\tabcolsep{2pt}
 \begin{tabular}{|c|c|c|c|c|c|c|c|c|c|}
  \hline
   Atom&{z}&1s&2s&2p&3s&3p&3d&4s&4p\\
 \hline\hline
Ga &31&   -361.309461 &      -43.037010 &      -37.727020 &       -4.576035 &       -2.951273 &       -0.264266 &       -0.165288 &       -0.002386 \\ 
 \hline
Ge &32&   -387.039855 &      -46.711685 &      -41.164308 &       -5.182760 &       -3.449483 &       -0.533749 &       -0.229337 &       -0.010542 \\ 
 \hline
As &33&   -413.704397 &      -50.583856 &      -44.796323 &       -5.856750 &       -4.011096 &       -0.860725 &       -0.293291 &       -0.022574 \\ 
 \hline
Se &33&   -441.297733 &      -54.647174 &      -48.616891 &       -6.590128 &       -4.628856 &       -1.240224 &       -0.358794 &       -0.037413 \\ 
 \hline
Br &35&   -469.815876 &      -58.896767 &      -52.621294 &       -7.377307 &       -5.297678 &       -1.668313 &       -0.426192 &       -0.054625 \\ 
 \hline
Kr &36&   -499.256211 &      -63.329305 &      -56.806329 &       -8.214637 &       -6.014298 &       -2.142323 &       -0.495638 &       -0.073991 \\ 
 \hline
 \end{tabular}
\end{center}

\noindent
\textbf{Fourth row:}
\begin{center}
\tiny\setlength\tabcolsep{2pt}
 \begin{tabular}{|c|c|c|c|c|c|c|c|c|c|c|c|}
 \hline
   Atom&{z}&1s&2s&2p&3s&3p&3d&4s&4p&5s&5p\\
 \hline\hline
Rb &37&   -529.827018 &      -68.150675 &      -61.378353 &       -9.306434 &       -6.983328 &       -2.867015 &       -0.760103 &       -0.271916 &       -0.008742 &-\\ 
 \hline
Sr &38&   -561.340511 &      -73.171957 &      -66.148672 &      -10.462839 &       -8.015158 &       -3.653051 &       -1.032665 &       -0.475893 &       -0.021586 &-\\ 
 \hline
Y &39&   -593.866153 &      -78.461974 &      -71.186183 &      -11.752114 &       -9.178245 &       -4.569112 &       -1.383317 &       -0.757307 &       -0.076589 &       -0.002707 \\ 
 \hline
\end{tabular}
\end{center}

\begin{center}
\tiny\setlength\tabcolsep{2pt}
 \begin{tabular}{|c|c|c|c|c|c|c|c|c|c|c|c|c|c|}
 \hline
   Atom&{z}&1s&2s&2p&3s&3p&3d&4s&4p&5s&5p&4d& $n$ (4d) \\
 \hline\hline
Zr &40&    -627.17364 &       -83.77963 &       -76.25111 &       -12.93681 &       -10.23529 &        -5.37710 &        -1.58204 &        -0.89248 &        -0.07367 &        -0.00048 &        -0.00048 &           0.3207  \\
 \hline
 \end{tabular}
\end{center}

\begin{center}
\tiny\setlength\tabcolsep{2pt}
 \begin{tabular}{|c|c|c|c|c|c|c|c|c|c|c|c|c|c|}
 \hline
   Atom&{z}&1s&2s&2p&3s&3p&3d&4s&4p&5s&6s&4d&$n$ (4d)\\
 \hline\hline
Nb &41&    -661.38533 &       -89.25420 &       -81.47185 &       -14.14588 &       -11.31538 &        -6.20667 &        -1.76423 &        -1.01284 &        -0.06267 &        -0.00014 &        -0.00014 &           0.5840  \\
 \hline
Mo &42&    -696.51265 &       -94.89727 &       -86.85999 &       -15.39104 &       -12.43032 &        -7.06960 &        -1.94236 &        -1.13016 &        -0.04957 &        -0.000002 &        -0.000002 &           0.7983  \\
 \hline
 \end{tabular}
\end{center}

\begin{center}
\tiny\setlength\tabcolsep{2pt}
 \begin{tabular}{|c|c|c|c|c|c|c|c|c|c|c|c|}
 \hline
   Atom&{z}&1s&2s&2p&3s&3p&3d&4s&4p&5s&4d \\
 \hline\hline  
Tc &43&   -732.565071 &     -100.718115 &      -92.424856 &      -16.681758 &      -13.589644 &       -7.975384 &       -2.126544 &       -1.254159 &       -0.044554 &       -0.009444 \\ 
 \hline
Ru &44&   -769.539351 &     -106.713582 &      -98.163269 &      -18.014957 &      -14.790336 &       -8.920979 &       -2.314092 &       -1.381847 &       -0.043203 &       -0.024185 \\ 
 \hline
Rh &45&   -807.43252 &     -112.88067 &     -104.07224 &      -19.3877  &      -16.02956 &       -9.90351 &       -2.50245 &       -1.51046 &       -0.04269 &       -0.04081 \\ 
 \hline
 \end{tabular}
\end{center}

\begin{center}
\tiny\setlength\tabcolsep{2pt}
 \begin{tabular}{|c|c|c|c|c|c|c|c|c|c|c|c|c|}
 \hline
   Atom&{z}&1s&2s&2p&3s&3p&3d&4s&4p&5s&4d&$n$ (4d)\\
 \hline\hline
Pd &46&    -846.21733 &      -119.19036 &      -110.12295 &       -20.77177 &       -17.27895 &       -10.89439 &        -2.66438 &        -1.61336 &        -0.03846 &        -0.03846 &           1.6655 \\
 \hline
Ag &47&    -885.91821 &      -125.66905 &      -116.34159 &       -22.19282 &       -18.56438 &       -11.91967 &        -2.82492 &        -1.71460 &        -0.03379 &        -0.03379 &           1.9293  \\
 \hline
 \end{tabular}
\end{center}

\begin{center}
\tiny\setlength\tabcolsep{2pt}
 \begin{tabular}{|c|c|c|c|c|c|c|c|c|c|c|c|c|}
 \hline
   Atom&{z}&1s&2s&2p&3s&3p&3d&4s&4p&4d&5s&5p\\
 \hline\hline
Cd &48&   -926.623485 &     -132.409803 &     -122.820764 &      -23.742846 &      -19.977847 &      -13.071777 &       -3.073669 &       -1.901671 &       -0.096713 &       -0.042861 &-\\ 
 \hline
In &49&   -968.415517 &     -139.493172 &     -129.641175 &      -25.501835 &      -21.599353 &      -14.430695 &       -3.487846 &       -2.251916 &       -0.310885 &       -0.131665 &       -0.002570 \\ 
 \hline
Sn &50&  -1011.130388 &     -146.755434 &     -136.639081 &      -27.305190 &      -23.264351 &      -15.832028 &       -3.900956 &       -2.599067 &       -0.517562 &       -0.181855 &       -0.010599 \\ 
 \hline
Sb &51&  -1054.799726 &     -154.228103 &     -143.846051 &      -29.184036 &      -25.004010 &      -17.307019 &       -4.343505 &       -2.973930 &       -0.749756 &       -0.230820 &       -0.021622 \\ 
 \hline
Te &52&  -1099.421697 &     -161.909103 &     -151.260063 &      -31.136080 &      -26.816070 &      -18.853453 &       -4.812921 &       -3.374212 &       -1.006149 &       -0.280095 &       -0.034651 \\ 
 \hline
I  &53&  -1144.994552 &     -169.796463 &     -158.879194 &      -33.159211 &      -28.698453 &      -20.469283 &       -5.307002 &       -3.797932 &       -1.285246 &       -0.330100 &       -0.049319 \\ 
 \hline
Xe  &54&  -1191.517037 &     -177.888737 &     -166.702039 &      -35.251891 &      -30.649647 &      -22.153023 &       -5.824212 &       -4.243727 &       -1.585928 &       -0.381026 &       -0.065446 \\ 
 \hline                                                                                                                                              
 \end{tabular}
\end{center}

\begin{remark} Our numerical simulations seem to show that for all $1\leq z\leq 54$, there are no unoccupied negative eigenvalues in the rHF ground states of neutral atoms.\end{remark}

\medskip

We end this section by the following figures, which back up the conjecture that rHF atomic densities are decreasing radial functions of the distance to the nucleus.  

\medskip

\begin{minipage}{\linewidth}
\centering
\includegraphics[width=0.3\textwidth,angle=270]{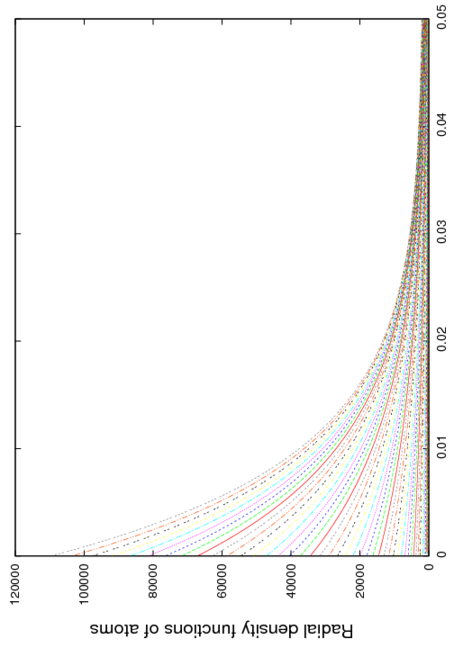}
\includegraphics[width=0.3\textwidth,angle=270]{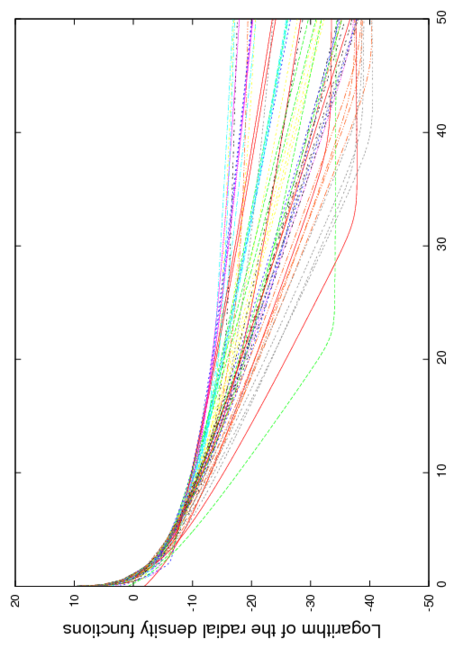}
\captionof{figure}{\scriptsize The left figure is the plot of the densities of all the atoms $1 \le z \le 54$ obtained with our code as a function of the distance to the nucleus, on the interval $[0,0.05]$. The right one is the plot of the logarithms of those densities on the interval $[0,50]$.}
\label{fig:1Ddensity}
\end{minipage}

\subsubsection{Occupied energy levels in the ${\rm X}\alpha$ model}
\label{sec:numXa}

The tables below provide the negative eigenvalues of the Kohn-Sham X$\alpha$ Hamiltonian (in Ha) for all the atoms of the first four rows of the periodic table . We observe that atoms $z$, with $23 \le z \le 28$ and $41 \le z \le 44$ have accidentally degenerate Fermi levels, the degeneracy occurring in all cases between an s-shell and a d-shell (4s-3d for $23 \le z \le 28$, 5s-4d for $41 \le z \le 44$) . All the results of this section are obtained for $L_e=30$ and $N_I$ increasing from $30$ to $75$ as $z$ increases.

\medskip

\noindent
\textbf{Hydrogen and helium:}
\begin{center}
\tiny
 \begin{tabular}{|c|c|c|}
\hline
  Atom&z& 1s \\
\hline\hline
H  & 1&     -0.194250 \\ 
 \hline
He & 2&     -0.516968 \\ 
 \hline
\end{tabular}
\end{center}

\noindent
\textbf{First row:}
\begin{center}
\tiny
\begin{tabular}{|c|c|c|c|c|}
\hline
 Atom&z&1s&2s&2p\\
\hline\hline
Li & 3&     -1.820596 &       -0.079032 &       -0.019804 \\ 
 \hline
Be & 4&     -3.793182 &       -0.170028 &       -0.045681 \\ 
 \hline
B  & 5&     -6.502185 &       -0.305377 &       -0.100041 \\ 
 \hline
C  & 6&     -9.884111 &       -0.457382 &       -0.157952 \\ 
 \hline
N  & 7&    -13.946008 &       -0.628841 &       -0.221004 \\ 
 \hline
O  & 8&    -18.690815 &       -0.820599 &       -0.289512 \\ 
 \hline
F  & 9&    -24.120075 &       -1.032963 &       -0.363534 \\ 
 \hline
Ne &10&    -30.234733 &       -1.266049 &       -0.443056 \\ 
 \hline
\end{tabular}
\end{center}

\noindent
\textbf{Second row:}
\begin{center}
\tiny
 \begin{tabular}{|c|c|c|c|c|c|c|}
  \hline
   Atom&z& 1s&2s&2p&3s&3p\\
 \hline\hline
Na &11&    -37.647581 &       -2.007737 &       -1.006028 &       -0.077016 &-\\ 
 \hline
Mg &12&    -45.897000 &       -2.845567 &       -1.661300 &       -0.142129 &-\\ 
 \hline
Al &13&    -55.080562 &       -3.877978 &       -2.507293 &       -0.251340 &       -0.071775 \\ 
 \hline
Si &14&    -65.107293 &       -5.017013 &       -3.456703 &       -0.359121 &       -0.117813 \\ 
 \hline
P  &15&    -75.982880 &       -6.269749 &       -4.516571 &       -0.470070 &       -0.166674 \\ 
 \hline
S  &16&    -87.709076 &       -7.638741 &       -5.689399 &       -0.585627 &       -0.218875 \\ 
 \hline
Cl &17&   -100.286615 &       -9.125221 &       -6.976378 &       -0.706438 &       -0.274567 \\ 
 \hline
Ar &18&   -113.715864 &      -10.729883 &       -8.378170 &       -0.832845 &       -0.333798 \\ 
 \hline
\end{tabular}
\end{center}

\noindent
\textbf{Third row:}
\begin{center}
\tiny\setlength\tabcolsep{2pt}
 \begin{tabular}{|c|c|c|c|c|c|c|c|c|}
 \hline
  Atom&{z} &   1s        &   2s       &    2p      &     3s    &    3p     &    4s      & 3d         \\
 \hline\hline
K  &19&   -128.330888 &      -12.775422 &      -10.219106 &       -1.233137 &       -0.646636 &       -0.064460 &-\\ 
 \hline
Ca &20&   -143.848557 &      -14.981138 &      -12.218289 &       -1.655845 &       -0.981391 &       -0.111359 &-\\ 
 \hline
Sc &21&   -160.10133 &      -17.14580 &      -14.17782 &       -1.94114 &       -1.18677 &       -0.12562 &       -0.08993 \\ 
 \hline
Ti &22&   -177.19446 &      -19.39840 &      -16.22419 &       -2.21070 &       -1.37630 &       -0.13516 &       -0.12742 \\ 
 \hline
\end{tabular}
\end{center}

\begin{center}
\tiny
 \begin{tabular}{|c|c|c|c|c|c|c|c|c|c|}
 \hline
  Atom&{z} &   1s     &   2s    &    2p   &     3s &    3p  &    4s  & 3d     &  $n$ (3d) \\
 \hline\hline
V  &23&    -195.11079 &       -21.72028 &       -18.33888 &        -2.44810 &        -1.53340 &        -0.13684 &        -0.13684 &          0.6393  \\
 \hline
Cr &24&    -213.87746 &       -24.14440 &       -20.55424 &        -2.68033 &        -1.68342 &        -0.13575 &        -0.13575 &          0.8873  \\
 \hline
Mn   &25 &    -233.50875 &       -26.68762 &       -22.88690 &        -2.92165 &        -1.83995 &        -0.13474 &        -0.13474 &          1.1278  \\
 \hline
Fe   &26 &    -254.00470 &       -29.35014 &       -25.33699 &        -3.17214 &        -2.00304 &        -0.13379 &        -0.13379 &          1.3622  \\
 \hline
Co   &27 &    -275.36535 &       -32.13212 &       -27.90468 &        -3.43191 &        -2.17274 &        -0.13292 &        -0.13292 &          1.5918  \\
 \hline
Ni &28&    -297.59075 &       -35.03372 &       -30.59009 &        -3.70102 &        -2.34907 &        -0.13212 &        -0.13212 &          1.8174  \\
 \hline
\end{tabular}
\end{center}

\begin{center}
\tiny\setlength\tabcolsep{2pt}
 \begin{tabular}{|c|c|c|c|c|c|c|c|c|c|}
  \hline
   Atom&{z}&    1s       &      2s    &     2p     &       3s  &     3p    &   3d      &    4s     &    4p     \\
 \hline\hline
Cu &29&   -320.711183 &      -38.088382 &      -33.426318 &       -4.010749 &       -2.562693 &       -0.157720 &       -0.138533 &-\\ 
 \hline
Zn &30&   -344.885966 &      -41.471174 &      -36.586685 &       -4.519851 &       -2.969457 &       -0.348234 &       -0.185366 &-\\ 
 \hline
Ga &31&   -370.087065 &      -45.140343 &      -40.030943 &       -5.188704 &       -3.532081 &       -0.685727 &       -0.290872 &       -0.070624 \\ 
 \hline
Ge &32&   -396.206872 &      -48.991790 &      -43.654803 &       -5.906101 &       -4.139819 &       -1.064181 &       -0.386783 &       -0.114696 \\ 
 \hline
As &33&   -423.248196 &      -53.026929 &      -47.459904 &       -6.673183 &       -4.794502 &       -1.487148 &       -0.481338 &       -0.158885 \\ 
 \hline
Se &33&   -451.209748 &      -57.243491 &      -51.444139 &       -7.487710 &       -5.494354 &       -1.953579 &       -0.576513 &       -0.20426 \\ 
 \hline
Br &35&   -480.090322 &      -61.639549 &      -55.605706 &       -8.347907 &       -6.237921 &       -2.462342 &       -0.673116 &       -0.251199 \\ 
 \hline
Kr &36&   -509.889039 &      -66.213681 &      -59.943283 &       -9.252538 &       -7.024197 &       -3.012574 &       -0.771572 &       -0.299874 \\ 
 \hline
\end{tabular}
\end{center}

\noindent
\textbf{Fourth row:}
\begin{center}
\tiny\setlength\tabcolsep{2pt}
 \begin{tabular}{|c|c|c|c|c|c|c|c|c|c|c|c|}
 \hline
  Atom&z  &   1s        &   2s       &    2p      &     3s     &    3p      &    3d      &   4s    &   4p  &   5s           & 4d\\
 \hline\hline
Rb &37&   -540.863861 &      -71.219637 &      -64.711316 &      -10.452293 &       -8.104015 &       -3.854833 &       -1.088064 &       -0.547366 &       -0.061487 &-\\ 
 \hline
Sr &38&   -572.774871 &      -76.418197 &      -69.670502 &      -11.708284 &       -9.238678 &       -4.750868 &       -1.407019 &       -0.798079 &       -0.102737 &-\\ 
 \hline
Y  &39&   -605.539841 &      -81.718973 &      -74.731216 &      -12.932519 &      -10.340292 &       -5.612293 &       -1.651693 &       -0.980422 &       -0.120721 &       -0.071919 \\ 
 \hline
Zr &40&   -639.200123 &      -87.167101 &      -79.938205 &      -14.171025 &      -11.455022 &       -6.485549 &       -1.873159 &       -1.141874 &       -0.131037 &       -0.111534 \\ 
 \hline
\end{tabular}
\end{center}

\begin{center}
\tiny\setlength\tabcolsep{2pt}
 \begin{tabular}{|c|c|c|c|c|c|c|c|c|c|c|c|c|}
 \hline
    Atom&z  &   1s  &   2s &    2p&   3s &    3p&  3d &   4s&   4p&  5s  & 4d   & $n$ (4d) \\
 \hline\hline
Nb &41&    -673.74149 &       -92.74707 &       -85.27606 &       -15.40918 &       -12.56830 &        -7.35588 &        -2.05942 &        -1.27048 &        -0.13172 &        -0.13172 &          0.6535  \\
 \hline
Mo &42&    -709.15136 &       -98.44597 &       -90.73190 &       -16.63439 &       -13.66757 &        -8.21062 &        -2.19877 &        -1.35425 &        -0.11937 &        -0.11937 &          0.9847  \\
 \hline
Tc &43&    -745.48044 &      -104.31826 &       -96.35989 &       -17.90004 &       -14.80629 &        -9.10349 &        -2.34006 &        -1.43939 &        -0.10617 &        -0.10617 &          1.2956  \\
 \hline
Ru &44&    -782.72787 &      -110.36286 &      -102.15896 &       -19.20531 &       -15.98365 &       -10.03361 &        -2.48279 &        -1.52544 &        -0.09183 &        -0.09183 &          1.5896  \\
 \hline
\end{tabular}
\end{center}

\begin{center}
\tiny\setlength\tabcolsep{2pt}
 \begin{tabular}{|c|c|c|c|c|c|c|c|c|c|c|c|c|}
 \hline
  Atom&z  &   1s        &   2s        &    2p       &     3s     &    3p      &    3d      &   4s      &   4p      &     4d    &    5s     & 5p\\
 \hline\hline
Rh &45&   -820.927173 &     -116.614569 &     -108.163817 &      -20.585170 &      -17.234646 &      -11.035987 &       -2.661143 &       -1.645733 &       -0.103288 &-&-\\ 
 \hline
Pd &46&   -860.048546 &     -123.041777 &     -114.343011 &      -22.008434 &      -18.528092 &      -12.079263 &       -2.845456 &       -1.771555 &       -0.118970 &-&-\\ 
 \hline
Ag &47&   -900.232540 &     -129.790427 &     -120.842024 &      -23.620128 &      -20.009041 &      -13.308869 &       -3.173860 &       -2.037653 &       -0.252103 &       -0.124136 &-\\ 
 \hline
Cd &48&   -941.381019 &     -136.759252 &     -127.559951 &      -25.317963 &      -21.575259 &      -14.622541 &       -3.543470 &       -2.343065 &       -0.420723 &       -0.167825 &-\\ 
 \hline
In &49&   -983.552576 &     -144.005647 &     -134.554225 &      -27.159345 &      -23.284171 &      -16.077676 &       -4.010922 &       -2.744597 &       -0.681578 &       -0.253924 &       -0.071162 \\ 
 \hline
Sn &50&  -1026.665599 &     -151.449408 &     -141.744613 &      -29.062993 &      -25.054553 &      -17.593291 &       -4.493043 &       -3.159222 &       -0.954355 &       -0.330583 &       -0.110212 \\ 
 \hline
Sb &51&  -1070.725180 &     -159.095276 &     -149.135914 &      -31.033521 &      -26.891049 &      -19.174056 &       -4.994724 &       -3.592188 &       -1.244953 &       -0.404626 &       -0.148390 \\ 
 \hline
Te &52&  -1115.731902 &     -166.943588 &     -156.728514 &      -33.071174 &      -28.793930 &      -20.820270 &       -5.516439 &       -4.044198 &       -1.554330 &       -0.477952 &       -0.186783 \\ 
 \hline
I  &53&  -1161.685673 &     -174.994060 &     -164.522166 &      -35.175601 &      -30.762871 &      -22.531629 &       -6.058048 &       -4.515264 &       -1.882595 &       -0.551382 &       -0.225814 \\ 
 \hline
Xe  &54&  -1208.586286 &     -183.246330 &     -172.516543 &      -37.346393 &      -32.797483 &      -24.307764 &       -6.619330 &       -5.005277 &       -2.229668 &       -0.625352 &       -0.265689 \\ 
 \hline
\end{tabular}
\end{center}

\medskip

We end this section by the following figures, which show that as in the ${\rm rHF}$ case, the ${\rm X}\alpha$ atomic densities seem to be decreasing radial functions of the distance to the nucleus.
  
\medskip

\begin{minipage}{\linewidth}
\centering
\includegraphics[width=0.3\textwidth,angle=270]{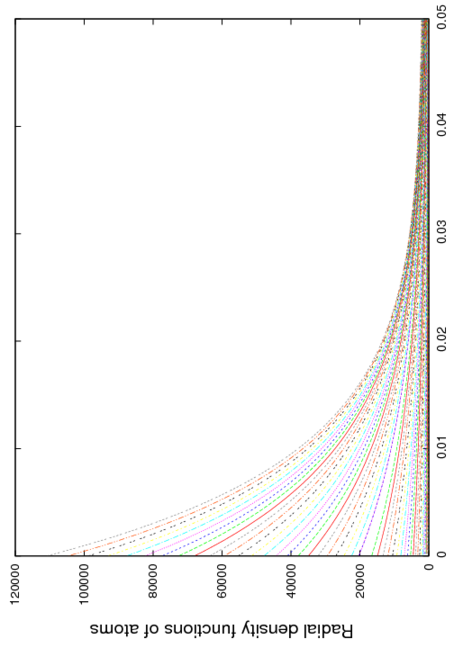}
\includegraphics[width=0.3\textwidth,angle=270]{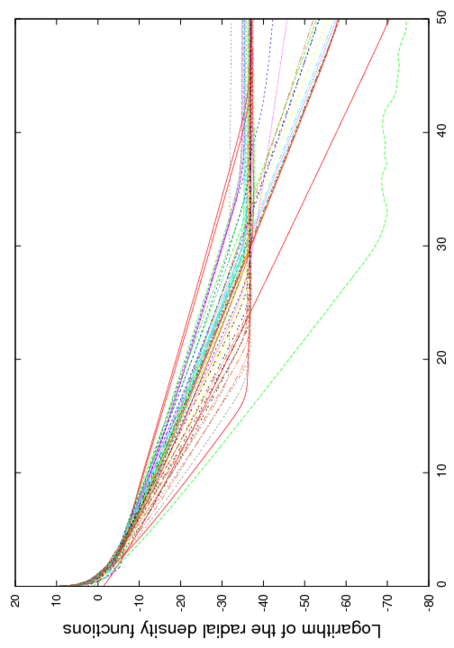}
\captionof{figure}{\scriptsize The left figure is the plot of the ${\rm X}\alpha$ densities of all the atoms $1 \le z \le 54$ obtained with our code as a function of the distance to the nucleus, on the interval $[0,0.05]$. The right one is the plot of the logarithms of those densities on the interval $[0,50]$.}
\label{fig:1Ddensity-xc=1}
\end{minipage}


\subsection{Perturbation by a uniform electric field (Stark effect)}
In this section, we consider atoms subjected to a uniform electric field, that is to an external potential $\beta W_{\rm Stark}$ with 
\begin{equation*}
W_{\rm Stark}(\br)=- e_{\bz} \cdot\br,
\end{equation*}
or, in spherical coordinates, 
$$
W_{\rm Stark}(r,\theta,\varphi)=-\sqrt{\frac{4\pi}{3}}r Y_1^0(\theta,\varphi).
$$
As already mentioned in Section~\ref{sec:perturbation},  $\widetilde\cI_{z,N}^{\rm rHF/LDA}(\beta W_{\rm Stark})=-\infty$ whenever $\beta \neq 0$, and the corresponding variational problem has no minimizer. However, one can find a minimizer $\gamma_h\in\cK_{N,h}$ to the approximated problem 
$\widetilde\cI_{z,N,h}^{\rm rHF/LDA}(\beta W_{\rm Stark})$. Hereafter we consider the carbon atom ($z=6$). Even though the cutoff $m_h$ is set equal to $6$, all the terms corresponding to a magnetic number $m>1$ are in fact equal to zero.

\medskip

The following figures are the plots in the $xy$-plane of the densities $\rho_h$ multiplied by $|\br|^2$  for the carbon atom ($z=6$) obtained for different values of $\beta$:
\vspace{-10mm}
\begin{figure}[h]
\begin{tabular}{cccc} \hspace{-10mm}
 \includegraphics[width=0.4\textwidth,angle=270]{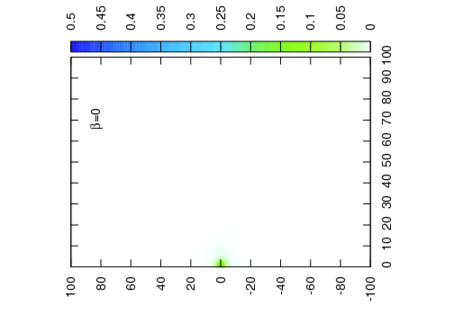} & \hspace{-5mm}
 \includegraphics[width=0.4\textwidth,angle=270]{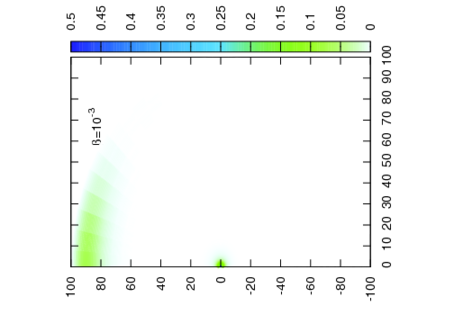} & \hspace{-5mm}
 \includegraphics[width=0.4\textwidth,angle=270]{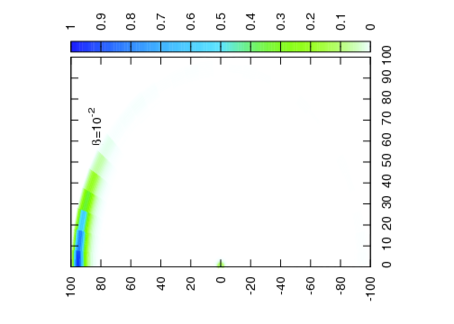} &\hspace{-5mm}
 \includegraphics[width=0.4\textwidth,angle=270]{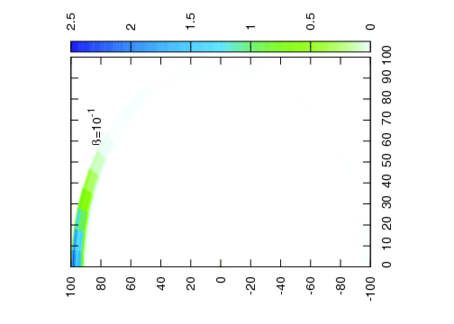}
\end{tabular}
\vspace{-10mm}
\caption{\scriptsize rHF case: the left figure is a plot of the density (multiplied by the function $(x^2+y^2)$) of an isolated  carbon atom. The other ones are plots of the densities  (multiplied by the function $(x^2+y^2)$) of the carbon atom subjected to a uniform external electric field, with coupling constants $\beta=10^{-3},10^{-2},10^{-1}$, respectively.}
\label{fig:densities_pert_z=6}
\end{figure}

\vspace{-10mm}
\begin{figure}[h]
\begin{tabular}{cccc} \hspace{-10mm}
 \includegraphics[width=0.4\textwidth,angle=270]{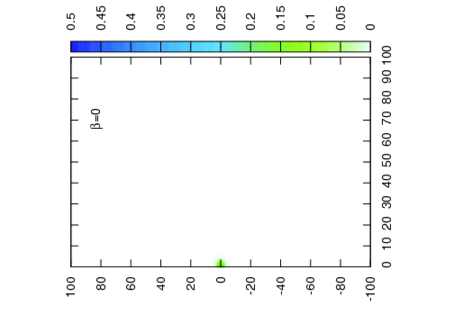} & \hspace{-5mm}
 \includegraphics[width=0.4\textwidth,angle=270]{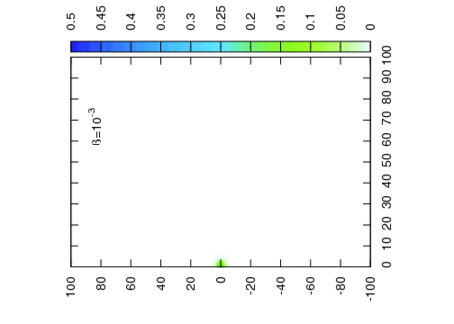} & \hspace{-5mm}
 \includegraphics[width=0.4\textwidth,angle=270]{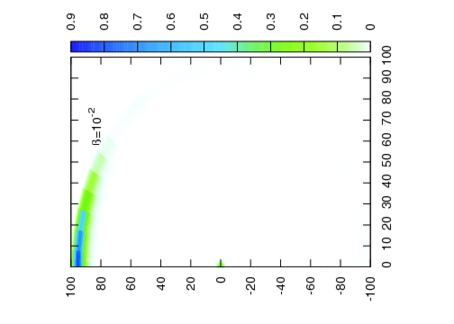} &\hspace{-5mm}
 \includegraphics[width=0.4\textwidth,angle=270]{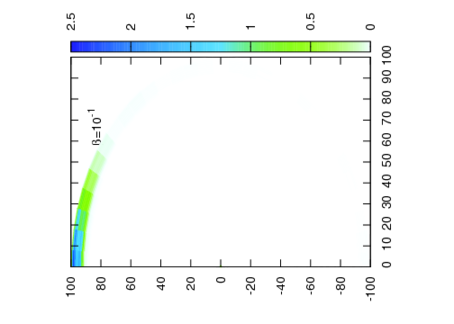}
\end{tabular}
\vspace{-10mm}
\caption{\scriptsize ${\rm X}\alpha$ case: The first figure is a plot of the density (multiplied by the function $(x^2+y^2)$) of an isolated carbon atom. The other ones are plots of the densities (multiplied by the function $(x^2+y^2)$) of the carbon atom subjected to a uniform external electric field, with coupling constants $\beta=10^{-3},10^{-2},10^{-1}$, respectively.}
\label{fig:xc_densities_pert_z=6}
\end{figure}



\medskip

For $\beta=10^{-2}$ and $\beta=10^{-1}$, we clearly see spurious boundary effects: part of the electronic cloud is localized in the region where the external potential takes highly negative values. This result is obviously not physical. On the other hand, for the X$\alpha$ model and for $\beta=10^{-3}$ we simply observe a polarization of the electronic cloud. The perturbation potential being not spherically symmetric, it breaks the symmetry of the density. This numerical solution can probably be interpreted as a (nonlinear) resonant state. We will come back to the analysis of this interesting case in a following work.


\medskip

Fig.~\ref{fig:polarization} shows the amount of electrons of the carbon atom which escape to infinity as a function of the coupling constant $\beta$ (for $L_e=100$ and $N_I=50$), in the rHF and ${\rm X}\alpha$ case.

\begin{minipage}{\linewidth}
\includegraphics[width=0.5\textwidth,angle=270]{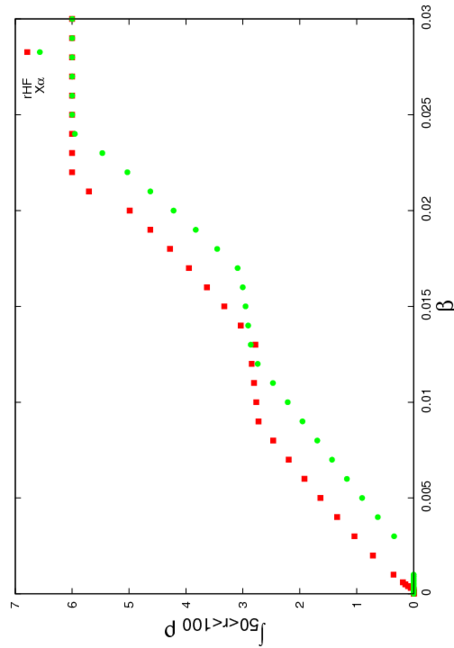}
\includegraphics[width=0.25\textwidth,angle=270]{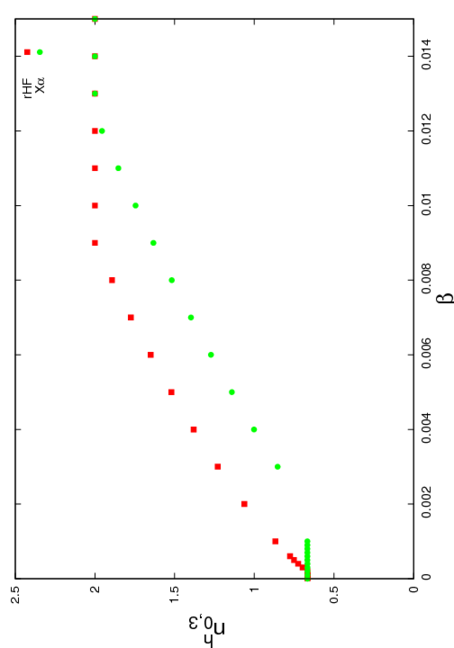}
\includegraphics[width=0.25\textwidth,angle=270]{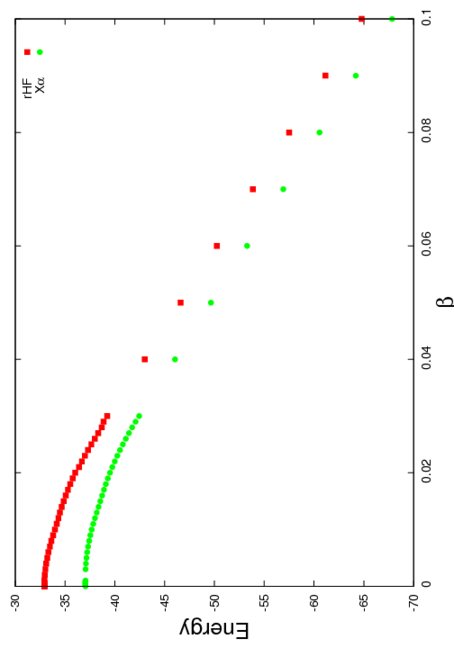}
\captionof{figure}{\scriptsize  The upper figure is the plot of the integral  on $B_{100}\setminus B_{50}$ of the density $\rho_h$, the lower left figure is the plot of the occupation number $n_{0,3}^h$, and the lower right one is the plot of the total energy, for $L_e=100$ and $N_I=50$ as a function of $\beta$ in the ${\rm rHF}$ and ${\rm X}\alpha$ cases.}
\label{fig:polarization}
\end{minipage}

\medskip

In general, the standard ODA is used to achieve convergence (see Section~\ref{sec:numerical-method}). However, for $\beta$ small (resp. large) enough, the occupation numbers are selected as follows:  $n_{0,1}^{[n]}=n_{0,2}^{[n]}=2$, $n_{0,3}=2(1-t_0)$ and $n_{1,1}^{[n]}=2-n_{0,3}=2 t_0$, $t_0$ being the minimizer of  
$$
t\mapsto \widetilde E_{6,6}^{\rm rHF/LDA} \left((1-t)\gamma^{[n]}_{0,*}+t\gamma^{[n]}_{1,*},\beta W\right),
$$
where 
$$
\gamma^{[n]}_{0,*}=2\sum_{1\le k\le 3}|\Phi_{0,k,h}\rangle\langle\Phi_{0,k,h}|\quad\mbox{and}\quad 
\gamma^{[n]}_{1,*}=2\sum_{1\le k\le 2}|\Phi_{0,k,h}\rangle\langle\Phi_{0,k,h}|+2|\Phi_{1,1,h}\rangle\langle\Phi_{1,1,h}|.
$$
This modification of ODA significantly increases the rate of convergence for $\beta$ small or large, but does not converge for all intermediate values of $\beta$.

\medskip

While $\widetilde\cI_{z,N}^{\rm rHF/LDA}(\beta W_{\rm Stark})=-\infty$ and the corresponding variational problem has no minimizer, the first-order perturbation $\gamma_{z,N,W_{\rm Stark}}^{(1),\rm rHF}$ of the ground state density matrix does exist (see Theorem~\ref{th:stark_effect}). If we consider the carbon atom, it can be expressed as a function of the unperturbed occupied Kohn-Sham orbitals and of their first-order perturbations. We indeed have
$$
\gamma_{6,6, W_{\rm Stark}}^{(1),\rm rHF}=\dps\sum_{\underset{\underset{i_1+i_2+i_3=1}{i_1\ge 0,i_2\ge 0,i_3\ge 0}}{(m,k)\in \cO_{6,6}}}  n_{m,k}^{(i_1)}|\Phi_{m,k}^{(i_2)}\rangle \langle \Phi_{m,k}^{(i_3)}|,
$$
where
$\cO_{6,6}=\{(0,1),(0,2),(0,3),(1,1)\}$, where $\epsilon_{m,k}^{(0)}$ is the $k$-th lowest eigenvalue of $H_{6,6}^{0,\rm rHF}$ in the subspace $\cH^m$, $\Phi_{m,k}^{(0)}$ an associated normalized eigenfunction and
$$
n_{0,1}^{(0)}=n_{0,2}^{(0)}=2, \quad n_{0,3}^{(0)}=\frac{2}{3} \quad\mbox{and}\quad n_{1,1}^{(0)}=4/3,
$$ 
while $\epsilon_{m,k}^{(1)}$, $\Phi_{m,k}^{(1)}$ and  $n_{m,k}^{(1)}$ satisfy the following self-consistent equation
\begin{align*}
& \left(H_{6,6}^{0,\rm rHF}-\epsilon_{m,k}^{(0)}\right)\Phi_{m,k}^{(1)}+\left(\rho^{(1)}\star|\cdot|^{-1}\right)\Phi_{m,k}^{(0)}+W_{\rm Stark}\Phi_{m,k}^{(0)}
=\epsilon_{m,k}^{(1)}\Phi_{m,k}^{(0)}, \\ \\
&\rho^{(1)}=\sum_{(m,k)\in\cO_{6,6}} \!\!\!\!\!\!\!\! 2\,n_{m,k}^{(0)}\Phi_{m,k}^{(0)}\Phi_{m,k}^{(1)} + n_{m,k}^{(1)}\Phi_{m,k}^{(0)}\Phi_{m,k}^{(0)},\\ \\
& \int_{\R^3} \Phi_{m,k}^{(1)}\Phi_{m,k}^{(0)}=0,\quad \mbox{and}\quad \sum_{(m,k)\in\cO_{6,6}}n_{m,k}^{(1)}=0.
\end{align*}

We denote by $\epsilon_{m,k,h}^{(0)}$, $\epsilon_{m,k,h}^{(1)}$, $\Phi_{m,k,h}^{(0)}$, $\Phi_{m,k,h}^{(1)}$ and $n_{m,k,h}^{(1)}$ the approximations of 
$\epsilon_{m,k}^{(0)}$, $\epsilon_{m,k}^{(1)}$, $\Phi_{m,k}^{(0)}$, $\Phi_{m,k}^{(1)}$ and $n_{m,k}^{(1)}$, respectively.
For each $(m,k)\in\cO_{6,6}$, define 
\begin{align*}
&\widetilde \epsilon_{m,k,h}^{(1)}(\beta):=\frac 1\beta (\epsilon_{m,k,h}(\beta)-\epsilon_{m,k,h}^{(0)}),\quad
\widetilde\Phi_{m,k,h}^{(1)}(\beta):=\frac 1\beta (\Phi_{m,k,h}(\beta)-\Phi_{m,k,h}^{(0)}),\quad\mbox{and}\quad\\
&\widetilde n_{m,k,h}^{(1)}(\beta):=\frac 1\beta (n_{m,k,h}(\beta)-n_{m,k}^{(0)}).
\end{align*}
Recall that, $\left(\Phi_{m,k,h}(\beta)\right)_{(m,k)\in\cO_{6,6}}$ (resp. $n_{m,k,h}^{(1)}(\beta)$) are the eigenfunctions (resp. eigenvalues) of the density matrix $\gamma_h$, the minimizer of the approximated problem $\widetilde\cI_{z,N,h}^{\rm rHF}(\beta W_{\rm Stark})$.

\medskip
Let $U^{m,k}$ and $\widetilde U^{m,k}(\beta)$ be such that
\begin{align*}
\Phi_{m,k,h}^{(0)}(r,\theta,\varphi)=\sum_{l=|m|}^{m_h}\left(\sum_{i=1}^{N_h}U_{i,l}^{m,k}(\beta)\cX_i(r)/r\right)Y_l^m(\theta,\varphi)\quad\mbox{and}\quad\\
\widetilde\Phi_{m,k,h}^{(1)}(\beta)(r,\theta,\varphi)=\sum_{l=|m|}^{m_h}\left(\sum_{i=1}^{N_h}\widetilde U_{i,l}^{m,k}(\beta)\cX_i(r)/r\right)Y_l^m(\theta,\varphi).
\end{align*}
To show that $\widetilde\Phi_{m,k,h}^{(1)}(\beta) \to \Phi_{m,k,h}^{(1)}$  when $\beta \to 0$, it is enough to show that for each $l\ge 0$

{ \small
\begin{align}
\left( \frac 12 A + \frac{l(l+1)}2 M_{-2} -z M_{-1}+NV_\mu-\epsilon^{(0)} M_0\right)\widetilde U_{.,l}(\beta)
-\frac{1}{\sqrt 3} C^{1,m} M_1 U_{.,l-1}-\frac{1}{\sqrt 3} C^{1,m} M_1 U_{.,l+1}\nonumber\\
+\sum_{l'=|m|}^{m_h}\sum_{l''=0}^{2m_h} C^{l,m}_{l',l''} ([Q_{l''}]^T\cdot F)\widetilde U_{.,l'}(\beta)
 + C^{l,m}_{l',l''} ([\widetilde Q_{l''}(\beta)]^T\cdot F) U_{.,l'}-\epsilon^{(1)}M_0 U_{.,l}\underset{\beta\rightarrow 0}{\rightarrow} 0.\label{eq:1st-pert}
\end{align}}
The index $(m,k)$ is omitted for simplicity and the vector $\widetilde Q_{l}(\beta)$ is the solution to the linear system
\begin{equation*}
\left(A^{\rm a}+l(l+1)M_{-2}^{\rm a}\right)\widetilde Q_l =4\pi F :\widetilde R_l,
\end{equation*} 
with
\begin{equation*}
\widetilde R_l := \dps\sum_{\underset{1\leq k\leq (m_h-|m|+1)\times N_h}{-m_h\leq m\leq m_h}}2 n_{m,k}^{(0)} \widetilde U^{m,k} C^{l,m} [\widetilde U^{m,k}]^T +n_{m,k}^{(1)} \widetilde U^{m,k} C^{l,m} [U^{m,k}]^T.
\end{equation*}

\medskip

\begin{figure}[h]
 \centering
 \includegraphics[width=0.3\textwidth,angle=270]{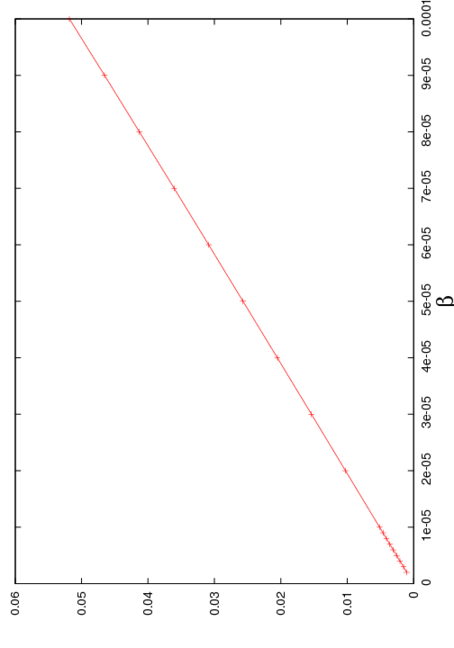}
 \caption{Plot of the function  $\beta \mapsto \max_{(m,k) \, | \; n_{m,k} > 0} \max_{l \ge |m|} \|V_{l,(m,k)}(\beta)\|_\infty$ where $V_{l,(m,k)}(\beta)$ is the vector in the left-hand side of (\ref{eq:1st-pert}).}
 \label{fig:1st-perturbation}
\end{figure}

Our numerical results show that, as expected by symmetry, $n_{m,k,h}^{(1)}=\epsilon_{m,k,h}^{(1)}=0$ for all $(m,k)\in\cO_{6,6}$, and that the left-hand side of (\ref{eq:1st-pert}) converges to zero linearly in $\beta$ (see Fig.\ref{fig:1st-perturbation}).

\section{Acknowledgements} The authors are grateful to Carlos Garc\'ia-Cervera and Vikram Gavini for valuable discussions.


\section*{Appendix: $\P4$ radial finite elements}\label{sec:FE}

In this appendix, we elaborate on the details of the calculation. 

\subsection*{A1. Basis functions}

\label{basis_form}

\noindent
We have chosen the following form functions to build the finite element matrices and tensors:
$$
z_1(t) = 1-t, \quad z_2(t) = t, \quad z_3(t) = 4t(1-t)=-4t^2+4t,
$$
$$
z_4(t) = \frac{128}3 t \left( \frac 12 -t \right)\left( \frac 34 -t \right)\left( 1 -t \right) = - \frac{128}3 \left( t^4 - \frac{9}4 t^3 + \frac{13}8 t^2 
- \frac 38 t \right),
$$
$$
z_5(t) = \frac{128}3 t \left(t -\frac 14 \right)\left( t- \frac 12  \right)\left( 1 -t \right)= - \frac{128}3 \left( t^4 - \frac{7}4 t^3 + \frac{7}8 t^2 
- \frac 18 t \right).
$$

\medskip

\noindent
Their derivatives are given by:
$$
z_1'(t) = -1, \quad z_2'(t) = 1, \quad z_3'(t) = -8t+4,
$$
$$
z_4'(t) = - \frac{128}3 \left( 4t^3 - \frac{27}4 t^2 + \frac{13}4 t - \frac 38  \right), \quad z_5'(t) = - \frac{128}3 \left( 4t^3 - \frac{21}4 t^2 + \frac{7}4 t 
- \frac 18  \right).
$$

\medskip

\noindent
Finite element basis:
\begin{itemize}
\item the 1D Schr\"odinger equation is solved on the finite interval $[0,L_e]$ with zero Dirichlet boundary conditions 
\item the interval $[0,L_{e}]$ is decomposed in $N_I$ intervals of positive lengths $h_1, \cdots h_{N_I}$. Let $0=r_1 < r_2 < \cdots < r_{N_I} < r_{N_I+1}=L_e$ be such that $h_{k}=r_{k+1}-r_k$;
\item we denote by
$$
V_h = \left\{ v \in C^0([0,L_e]) \mbox{ s.t. } v|_{[r_k,r_{k+1}]} \in \P_4, \quad v(0)=v(L_e)=0 \right\}
$$
the $\P_4$ finite element space associated with the so-defined mesh. We have
$$
\mbox{dim}(V_h) = 4N_I-1;
$$
\item we then set for all $1 \le k \le N_I$ and $1 \le j \le 5$,
$$
p^k_j(r)=z_j\left( \frac{r-r_k}{h_k} \right) 
$$
so that $p^k_j(r_k+th_k)=z_j(t)$, and define the basis $(\chi_1,\cdots, \chi_{4N_I-1})$ of $V_h$ as follows:
$$
\chi_1(r)=p^1_3(r) \1_{[r_1,r_2]}, \quad \chi_2(r)=p^1_4(r)\1_{[r_1,r_2]}, \quad \chi_3(r)=p^1_5(r)\1_{[r_1,r_2]},
$$
and for all $2 \le k \le N_I$, 
$$
\left\{ \begin{array}{l}\dps 
\chi_{4k-4}(r)=p^{k-1}_2(r)\1_{[r_{k-1},r_k]}+p^k_1(r)\1_{[r_{k},r_{k+1}]} \\
\chi_{4k-3}(r)=p^k_3(r)\1_{[r_{k},r_{k+1}]} \\
\chi_{4k-2}(r)=p^k_4(r)\1_{[r_{k},r_{k+1}]} \\
\chi_{4k-1}(r)=p^k_5(r)\1_{[r_{k},r_{k+1}]} \end{array}\right.
$$
\item when considering an atom with Fermi level very close or possible equal to zero, an extra basis function of the form 
$$
\chi_{4 N_I}(r)=p^{N_I}_2(r)\1_{[r_{N_I},r_{N_I+1}]}+\frac{r_{N_I+1}}{r}\1_{[r_{N_I+1},\infty[}
$$
is added to the space $V_h$. Its derivative is equal to
$$
\chi_{4 N_I}'(r)=\frac{1}{h_{N_I}}\1_{[r_{N_I},r_{N_I+1}[}-\frac{r_{N_I+1}}{r^2}\1_{]r_{N_I+1},\infty[.}
$$
\end{itemize}

\subsection*{A2. Assembling the matrices}\label{sec:assem-mat}
Let $\Lambda$ be the bijective mapping from  $\{0,1,2,3,4\}$ to $\{1,2,3,4,5\}$ defined by 
$$\Lambda(0)=2,\quad \Lambda(1)=5,\quad \Lambda(2)=4,\quad \Lambda(3)=3,\quad\mbox{and}\quad\Lambda(4)=1.$$  
Recall that the density is equal to 
$$
\rho_h(r,\theta)=\dps\sum_{l=0}^{2m_h}\dps\sum_{i,j=1}^{N_h}[R_l]_{i,j} \frac{\cX_i(r)}{r}\frac{\cX_j(r)}{r} Y_l^0(\theta).
$$
Using the finite element basis defined above, one gets that $\rho_h(r,\theta)$ is equal to
$$
  \left|\begin{array}{rl}\dps\sum_{l=0}^{2m_h}\dps\sum_{i,j=0}^{3}[R_l]_{4-i,4-j}\frac{p^1_{\Lambda(i)}(r)}{r}\frac{p^1_{\Lambda(j)}(r)}{r}Y_l^0(\theta)
                               &\quad \mbox{if}\quad r\in(r_1,r_2)\\ \\
                          \dps\sum_{l=0}^{2m_h}\dps\sum_{i,j=0}^{4}[R_l]_{4k-i,4k-j}\frac{p^k_{\Lambda(i)}(r)}{r}\frac{p^k_{\Lambda(j)}(r)}{r}Y_l^0(\theta)
                               &\quad \mbox{if}\quad r\in(r_k,r_{k+1}),\quad 1<k<N_I\\ \\ 
                          \dps\sum_{l=0}^{2m_h}\dps\sum_{i,j={\kappa}}^{4}[R_l]_{4N_I-i,4N_I-j}\frac{p^{N_I}_{\Lambda(i)}(r)}{r}\frac{p^{N_I}_{\Lambda(j)}(r)}{r}Y_l^0(\theta)
                               &\quad \mbox{if}\quad r\in(r_{N_I},r_{N_I+1}),\end{array}\right.
$$
where $\kappa$ equal to $1$ (resp. $0$) if the discretization space $V_h$ (resp. $V_h\cup\{\chi_{4N_I}\}$) is considered.
In particular, for $0< t_{p,\rm r} < 1$ and $-1<t_{q,\theta}<1$, we have that $(t_{p,\rm r}h_k+r_k)^2\rho(t_{p,\rm r}h_k+r_k,\arccos(t_{q,\theta}))$ is equal to
\begin{eqnarray}
  \left|\begin{array}{rl}\dps\sum_{l=0}^{2m_h}\dps\sum_{i,j=0}^{3}[R_l]_{4-i,4-j}z_{\Lambda(i)}(t_{p,\rm r})z_{\Lambda(j)}(t_{p,\rm r})\sqrt{\frac{2l+1}{4\pi}}P_l(t_{q,\theta})
                               &\quad \mbox{if}\quad k=1\\ \\
  \dps\sum_{l=0}^{2m_h}\dps\sum_{i,j=0}^{4}[R_l]_{4k-i,4k-j}z_{\Lambda(i)}(t_{p,\rm r})z_{\Lambda(j)}(t_{p,\rm r})\sqrt{\frac{2l+1}{4\pi}}P_l(t_{q,\theta})
                               &\quad \mbox{if}\quad 1<k<N_I\\ \\
  \dps\sum_{l=0}^{2m_h}\dps\sum_{i,j={\kappa}}^{4}[R_l]_{4N_I-i,4N_I-j}z_{\Lambda(i)}(t_{p,\rm r})z_{\Lambda(j)}(t_{p,\rm r})\sqrt{\frac{2l+1}{4\pi}}P_l(t_{q,\theta})
                               &\quad \mbox{if}\quad k=N_I,\end{array}\right.\label{density-at-pt}
\end{eqnarray}
\medskip
where $P_l$ are the Legendre polynomials, which can be calculated using the recurrence relation
\begin{eqnarray*}
 P_n(x)=\frac{2n-1}{n}xP_{n-1}(x)-\frac{n-1}{n}P_{n-2}(x),\quad n\geq 2,
\end{eqnarray*}
with $P_0(x)=1$ and $P_1(x)=x$. Note that system (\ref{density-at-pt}) is used only to calculate the exchange-correlation potential, thus there is no need to define the density at radius greater than $r_{N_{I+1}}$.

\medskip

For $\mu(r\be)= \frac{\eta^2}{4\pi}\frac{e^{-\eta r}}{r}$, then
$$
[V^{\rm H}(\mu)] (r\be)=\frac{1}{r}\left(1-e^{-\eta r}\right).
$$ 
Thus the vector $G$ in (\ref{vectorG}) has one of the following form depending on the discretization space considered
$$
G=\left[g^1_3,g^1_4,g^1_5,\cdots,g^{k-1}_2+g^k_1,g^k_3,g^k_4,g^k_5,\cdots,g^{N_I-1}_2+g^{N_I}_1,g^{N_I}_3,g^{N_I}_4,g^{N_I}_5\right]^T,
$$
or
$$
G=\left[g^1_3,g^1_4,g^1_5,\cdots,g^{k-1}_2+g^k_1,g^k_3,g^k_4,g^k_5,\cdots,g^{N_I-1}_2+g^{N_I}_1,g^{N_I}_3,g^{N_I}_4,g^{N_I}_5,g^{N_I}_2+g_\infty\right]^T,
$$
where 
$$
 g_i^k=\frac{\eta^2}{4\pi}h_k e^{-\eta r_k}\int_0^1 e^{-\eta t h_k}z_i(t) \, dt 
$$
and
$$
g_\infty=\frac{\eta^2}{4\pi}r_{N_{I+1}}\int_{r_{N_{I+1}}}^\infty \frac{e^{-\eta r}}{r}dr=\frac{\eta^2}{4\pi}r_{N_{I+1}}\int_0^1 \frac{e^{\frac{-\eta r_{N_{I+1}}}{t}}}{t}\, dt.
$$
We denote by
\begin{equation}\label{def:mat-coulumb}
 [\hat\cH _{l,l'}]=\sum_{l''=0}^{2m_h} C^{l,m}_{l',l''} \left([Q_{l''}]^T\cdot F\right),
\end{equation}
where $C^{l,m}$, $Q_l$ and $F$ are defined by (\ref{eq:matrixClm}), (\ref{eq:Poisson_l_2}) and (\ref{tensorF}), respectively.

\medskip

All the matrices $A$, $M_{-2}$, $M_{-1}$, $M_0$, $M_1$, $V_\mu$, $[V_{\rm xc}^l]$ and $[\hat\cH _{l,l'}]$ defined in (\ref{matrices}), (\ref{def:V_mu}), 
(\ref{eq:Vxcrhoh}) and (\ref{def:mat-coulumb}), when considering the discretization space $V_h$, are symmetric and have the same pattern:

 \begin{figure}[h]
 \centering
 \includegraphics[width=0.7\textwidth]{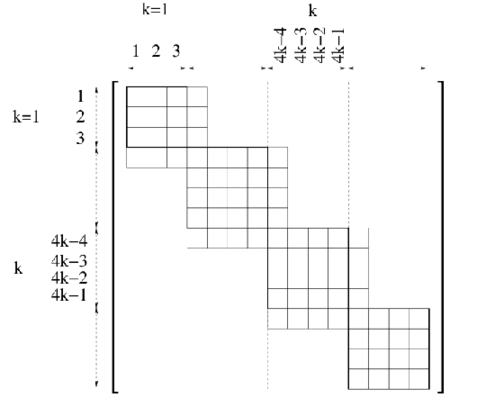}
 \end{figure}

\newpage

Their entries can be computed using elementary assembling matrices:
\begin{itemize}
\item diagonal blocks: for any $1 \le k \le N_I$, 
$$
\begin{array}{llll}
Y_{4k-4,4k-4}=y^{k-1}_{22}+y^k_{11} & Y_{4k-4,4k-3}=y^k_{13} & Y_{4k-4,4k-2}=y^k_{14} & Y_{4k-4,4k-1}=y^k_{15} \\
Y_{4k-3,4k-4}=y^k_{31} & Y_{4k-3,4k-3}=y^k_{33} & Y_{4k-3,4k-2}=y^k_{34} & Y_{4k-3,4k-1}=y^k_{35} \\
Y_{4k-2,4k-4}=y^k_{41} & Y_{4k-2,4k-3}= y^k_{43}& Y_{4k-2,4k-2}=y^k_{44} & Y_{4k-2,4k-1}=y^k_{45} \\
Y_{4k-1,4k-4}=y^k_{51} & Y_{4k-1,4k-3}=y^k_{53} & Y_{4k-1,4k-2}=y^k_{54} & Y_{4k-1,4k-1}= y^k_{55}
\end{array}
$$
\item off-diagonal blocks: for $1\leq k\leq N_I-1$
$$
Y_{4k-4,4k}=y^k_{12}, \; Y_{4k-3,4k}=y^k_{23}, \; Y_{4k-2,4k}=y^k_{24}, \; Y_{4k-1,4k}=y^k_{25},
$$
$$
Y_{4k,4k-4}=y^k_{21}, \; Y_{4k,4k-3}=y^k_{32}, \; Y_{4k,4k-2}=y^k_{42}, \; Y_{4k,4k-1}=y^k_{52}.
$$
\end{itemize}
When $\chi_{4N_I}$ is added to the discretization space, an extra row and an extra column must be added to each of the above matrices. The non-zero additional entries are:
$$
Y_{4N_I,4N_I}=y^{N_I}_{22}+y_\infty,
$$
$$
Y_{4N_I-4,4N_I}=y^{N_I}_{12}, \; Y_{4N_I-3,4N_I}=y^{N_I}_{23}, \; Y_{4N_I-2,4N_I}=y^{N_I}_{24}, \; Y_{4N_I-1,4N_I}=y^{N_I}_{25},
$$
$$
Y_{4N_I,4N_I-4}=y^{N_I}_{21}, \; Y_{4N_I,4N_I-3}=y^{N_I}_{32}, \; Y_{4N_I,4N_I-2}=y^{N_I}_{42}, \; Y_{4N_I,4N_I-1}=y^{N_I}_{52}.
$$

\noindent
The $y^k_{ij}$'s are the entries of the elementary assembling matrices. The latter are defined for the matrices $A$, $M_{-2}$, $M_{-1}$, $M_0$, $M_1$, $V_\mu$, 
 $[V_{\rm xc}^l]$ and $[\hat\cH _{l,l'}]$ as follows:

\begin{eqnarray*}
a^k_{ij} &=& \int_{r_k}^{r_{k+1}} {p^k_i}' {p^k_j}' = h_k^{-1} \int_0^1 z_i'z_j' =  h_k^{-1} \alpha_{ij} \\
(m_{-2})^k_{ij} &=& \int_{r_k}^{r_{k+1}} \frac{p^k_i(r) p^k_j(r)}{r^2} \, dr = h_k \int_0^1 \frac{z_i(t)z_j(t)}{(r_k+th_k)^2} \, dt\\
         &=& \left| \begin{array}{ll} \dps h_k r_k^{-2} \int_0^1 \frac{z_i(t)z_j(t)}{(1+th_k/r_k)^2} \, dt & \mbox{ if } k \ge 2 \\
              \dps h_1^{-1} \int_0^1 \frac{z_i(t)z_j(t)}{t^2} \, dt  & \mbox{ if } k =1 \end{array} \right. 
\\
(m_{-1})^k_{ij} &=& \int_{r_k}^{r_{k+1}} \frac{p^k_i(r) p^k_j(r)}{r} \, dr = h_k \int_0^1 \frac{z_i(t)z_j(t)}{r_k+th_k} \, dt\\ 
         &=& \left| \begin{array}{ll} \dps h_k r_k^{-1} \int_0^1 \frac{z_i(t)z_j(t)}{1+th_k/r_k} \, dt & \mbox{ if } k \ge 2 \\
\dps  \int_0^1 \frac{z_i(t)z_j(t)}{t} \, dt  & \mbox{ if } k =1 \end{array} \right. 
\\
(m_0)^k_{ij} &=& \int_{r_k}^{r_{k+1}} p^k_i(r) p^k_j(r)dr = h_k \int_0^1 z_i(t)z_j(t)dt =  h_k \, \nu_{ij} \\
(m_{1})^k_{ij} &=& \int_{r_k}^{r_{k+1}} r p^k_i(r) p^k_j(r)dr = h_k^2 \int_0^1 tz_i(t)z_j(t)dt+h_kr_k\,\nu_{ij}= h_k^2\beta_{ij}+h_kr_k\,\nu_{ij}\\
(v_\mu)_{ij}^k&=& (m_{-1})_{ij}^k- h_k e^{-\eta r_k}\int_0^1\frac{e^{-\eta t h_k}}{r_k+th_k}z_i(t)z_j(t)\, dt\\
            &=& \left| \begin{array}{ll} \dps  (m_{-1})^k- h_kr_k^{-1} e^{-\eta r_k}\int_0^1\frac{e^{-\eta t h_k}}{1+th_k/r_k}z_i(t)z_j(t)\, dt& \mbox{ if } k \ge 2 \\
                 \dps   (m_{-1})^k-  e^{-\eta r_k}\int_0^1\frac{e^{-\eta t h_k}}{t}z_i(t)z_j(t)\, dt  & \mbox{ if } k =1 \end{array} \right. \\
\end{eqnarray*}

\begin{eqnarray*}
(v_{\rm xc}^l)_{ij}^k
&=&c_{\rm xc}h_k\sqrt{\frac{2l+1}{4\pi}} \dps\sum_{p=1}^{N_{g,\rm r}}\dps\sum_{q=1}^{N_{g,\theta}}\omega_p\omega_q' 
                            \left(\rho(t_{p,\rm r}h_k+r_k,\arccos(t_{q,\theta}))\right)^{\frac 13}P_l(t_{q,\theta})z_i(t_{p,\rm r})z_j(t_{p,\rm r})\\
&=&c_{\rm xc}\left|\begin{array}{rl}h_kr_k^{-1}\sqrt{\frac{2l+1}{4\pi}} \dps\sum_{p=1}^{N_{g,\rm r}}\dps\sum_{q=1}^{N_{g,\theta}}\omega_p\omega_q' 
                            \left((t_{p,\rm r}h_k+r_k)^2\rho(t_{p,\rm r}h_k+r_k,\arccos(t_{q,\theta}))\right)^{\frac 13}\\
    P_l(t_{q,\theta})\frac{z_i(t_{p,\rm r})z_j(t_{p,\rm r})}{t_{p,\rm r}h_k/r_k+1}(t_{p,\rm r}h_k+r_k)^{\frac 13}\quad \mbox{ if } k \ge 2 \\
    \sqrt{\frac{2l+1}{4\pi}} \dps\sum_{p=1}^{N_{g,\rm r}}\dps\sum_{q=1}^{N_{g,\theta}}\omega_p\omega_q' \left((t_{p,\rm r}h_k)^2\rho(t_{p,\rm r}h_k+r_k,\arccos(t_{q,\theta}))\right)^{\frac 13}\\
    P_l(t_{q,\theta})\frac{z_i(t_{p,\rm r})z_j(t_{p,\rm r})}{t_{p,\rm r}}(t_{p,\rm r}h_k)^{\frac 13}\quad \mbox{ if } k = 1             \end{array}\right.
\end{eqnarray*}
in the ${\rm X}\alpha$-case, that is for $v_{\rm xc}(\rho)=-\left(\frac 3\pi\right)^{\frac 13} \rho^{\frac 13}$,
\begin{eqnarray*}
(\hat h_{l,l'})_{ij}^k&=&\left| \begin{array}{ll}\dps\sum_{l''=0}^{2m_h}\sum_{n=1}^{3}c_{l,l',l''}^{m}f_{ij\Lambda(n)}^k Q_{l'',4-n}& \mbox{ if } k =1\\
                                               \dps\sum_{l''=0}^{2m_h}\sum_{n=0}^{4}c_{l,l',l''}^{m}f_{ij\Lambda(n)}^k Q_{l'',4k-n}& \mbox{ if } 1< k<N_I \\ 
                                                \dps\sum_{l''=0}^{2m_h}\sum_{n={\kappa}}^{4}c_{l,l',l''}^{m}f_{ij\Lambda(n)}^k Q_{l'',4N_I-n}& \mbox{ if }  k=N_I, 
                                                 \end{array} \right.
\end{eqnarray*}
\medskip
where
$$
f_{ijn}^k=\int_{r_k}^{r_{k+1}}\frac{p_i^k(r)p_j^k(r)p_n^k(r)}{r}dr=h_k\int_0^1\frac{z_i(t)z_j(t)z_n(t)}{(th_k+r_k)}dt,
$$
and
$$
c_{\rm xc}=-\sqrt{\pi}\left(\frac 3\pi\right)^{\frac 13}.
$$
Note that  $\rho(t_{p,\rm r}h_k+r_k,\arccos(t_{q,\theta}))$ is calculated with the help of (\ref{density-at-pt}). The $y_\infty$ are defined for the matrices $A$, $M_{-2}$, $M_{-1}$, $M_{0}$, $V_\mu$ and $[\hat H_{l,l']}$ as follows:
$$
 a_\infty       =\int_{r_{N_{I+1}}}^\infty \chi_{4N_I}'\chi_{4N_I}'=\int_{r_{N_{I+1}}}^\infty \frac{(r_{N_{I+1}})^2}{r^4}dr=\frac{1}{3 r_{N_{I+1}}},
 $$
 $$
 (m_{-2})_\infty=\int_{r_{N_{I+1}}}^\infty \frac{\chi_{4N_I}(r)\chi_{4N_I}(r)}{r^2}dr=\int_{r_{N_{I+1}}}^\infty \frac{(r_{N_{I+1}})^2}{r^4}dr=\frac{1}{3 r_{N_{I+1}}},
 $$
 $$
 (m_{-1})_\infty=\int_{r_{N_{I+1}}}^\infty \frac{\chi_{4N_I}(r)\chi_{4N_I}(r)}{r}dr=\int_{r_{N_{I+1}}}^\infty \frac{(r_{N_{I+1}})^2}{r^3}dr=\frac12,
 $$
 $$
 (m_0)_\infty   =\int_{r_{N_{I+1}}}^\infty \chi_{4N_I}\chi_{4N_I}=\int_{r_{N_{I+1}}}^\infty \frac{(r_{N_{I+1}})^2}{r^2}dr=r_{N_{I+1}},
 $$
 \begin{align*}
 (v_\mu)_\infty &=\int_{r_{N_{I+1}}}^\infty \frac{1}{r}\left(1-e^{-\eta r}\right)\chi_{4N_I}(r)\chi_{4N_I}(r)dr=(m_{-1})_\infty-(r_{N_{I+1}})^2\int_{r_{N_{I+1}}}^\infty \frac{ e^{-\eta r}}{r^3}dr  \\ &=(m_{-1})_\infty+\int_0^1  e^{-\frac{\eta r_{N_{I+1}}}{t}} t\, dt,
\end{align*}
and
\begin{eqnarray*}
 (\hat h_{l,l'})_\infty=\dps\sum_{l''=0}^{2m_h}c_{l,l',l''}^{m}f_\infty Q_{l'',4N_I},
\end{eqnarray*}
where $f_\infty=\int_{r_{N_{I+1}}}^\infty \frac{\chi_{4N_I}^3(r)}{r}dr=\int_{r_{N_{I+1}}}^\infty \frac{(r_{N_{I+1}})^3}{r^4}dr=\frac 13$.

\medskip

In addition to assembling the matrices, we need to deal with the following  term
$$
\dps\sum_{i,j=1}^{N_h}F_{ijn}[R_l]_{ij}
$$
in order to calculate the right-hand side of (\ref{eq:Poisson_l_2}).
Let $k_n=1+\mbox{int}(\frac n4)$ and $q_n=4-(n \, \mbox{mod} \, 4)$, so that $n=4k_n-q_n$. Then the entries $\dps\sum_{i,j=1}^{N_h}F_{ijn}[R_l]_{ij}$ of the vector $F : R_l$ are computed as follows
$$
\left|\begin{array}{ll} 
\dps\sum_{i,j=0}^3 f_{\Lambda(i)\Lambda(j)\Lambda(q_n)}^{1}[R_l]_{4-i,4-j}\quad \mbox{if}\quad k_n=1 \\ \\
\dps\sum_{i,j=0}^4 f_{\Lambda(i)\Lambda(j)\Lambda(q_n)}^{k_n}[R_l]_{4k_n-i,4k_n-j}
         \quad \mbox{if}\quad q_n\neq 4 \quad \mbox{and}\quad 1<k_n<N_I\\ \\
\dps\sum_{i,j={\kappa}}^4 f_{\Lambda(i)\Lambda(j)\Lambda(q_n)}^{N_I}[R_l]_{4N_I-i,4N_I-j} 
         \quad \mbox{if}\quad q_n\neq 4 \quad \mbox{and}\quad k_n=N_I\\ \\
\begin{array}{l}\dps\sum_{i,j=0}^3 f_{2\Lambda(i)\Lambda(j)}^{1}[R_l]_{4-i,4-j}+ \dps\sum_{i,j=0}^4 f_{1\Lambda(i)\Lambda(j)}^{2}[R_l]_{8-i,8-j} \\  
            \quad\quad\quad\quad\quad\quad\quad\quad\quad\quad\quad\quad\quad\quad\quad\quad\quad\quad \mbox{if}\quad q_n=4\quad \mbox{and}\quad k=2\end{array}\\ \\
\begin{array}{c}\dps\sum_{i,j=0}^4\left[ f_{2\Lambda(i)\Lambda(j)}^{k_n-1}[R_l]_{4k_n-4-i,4k_n-4-j}\right.\\ \\
            \left.+f_{1\Lambda(i)\Lambda(j)}^{k_n}[R_l]_{4k_n-i,4k_n-j}\right]\end{array}\quad \mbox{if}\quad q_n=4\quad\mbox{and}\quad 2<k<N_I\\ \\
\begin{array}{c}\dps\sum_{i,j=0}^4 f_{2\Lambda(i)\Lambda(j)}^{N_I-1}[R_l]_{4N_I-4-i,4N_I-4-j}\\ \\
                +\dps\sum_{i,j={\kappa}}^4 f_{1\Lambda(i)\Lambda(j)}^{N_I}[R_l]_{4N_I-i,4N_I-j}\end{array}
            \quad \mbox{if}\quad q_n=4\quad\mbox{and}\quad k=N_I. \end{array}\right.                                                                                                                                                                                                                                                                                                                                  
$$

When the base $\chi_{4N_I}$ is added, the last entry of the vector $F : R_l$ is equal to
$$
\left(F : R_l\right)_{N_I}=\dps\sum_{i,j={\kappa}}^4 f_{2\Lambda(i)\Lambda(j)}^{N_I}[R_l]_{4N_I-i,4N_I-j}+ f_\infty [R_l]_{4N_I,4N_I}.
$$

We end this section by providing the values of $\alpha_{ij}$, $\beta_{ij}$ and $ \nu_{ij}$, for $1\leq i,j\leq 5$,
$$
\begin{array}{lllll}
\alpha_{11}= 1 & \alpha_{12}= -1 & \alpha_{13}= 0 & \alpha_{14}= 0 & \alpha_{15}= 0 \\
\alpha_{21}= -1 & \alpha_{22}= 1 & \alpha_{23}= 0 & \alpha_{24}= 0 & \alpha_{25}= 0 \\
\alpha_{31}= 0 & \alpha_{32}= 0 & \alpha_{33}= 16/3 & \alpha_{34}=128/45 & \alpha_{35}=128/45 \\
\alpha_{41}= 0 & \alpha_{42}= 0 & \alpha_{43}=128/45 & \alpha_{44}= 3328/189 & \alpha_{45}= 5888/945 \\
\alpha_{51}= 0 & \alpha_{52}= 0 & \alpha_{53}=128/45 & \alpha_{54}=5888/945 & \alpha_{55}= 3328/189
\end{array}
$$
$$
\begin{array}{lllll}
\nu_{11}= 1/3 & \nu_{12}=1/6 & \nu_{13}= 1/3 & \nu_{14}= 4/15 & \nu_{15}=4/45  \\
\nu_{21}= 1/6 & \nu_{22}=1/3 & \nu_{23}= 1/3 & \nu_{24}= 4/45 & \nu_{25}=4/15 \\
\nu_{31}= 1/3 & \nu_{32}=1/3 & \nu_{33}= 8/15 & \nu_{34}= 64/315 & \nu_{35}=64/315 \\
\nu_{41}= 4/15 & \nu_{42}= 4/45 & \nu_{43}= 64/315 & \nu_{44}=128/405 & \nu_{45}= 128/2835 \\
\nu_{51}=4/45  & \nu_{52}=4/15 & \nu_{53}= 64/315 & \nu_{54}=128/2835 & \nu_{55}=128/405
\end{array}
$$ 
$$
\begin{array}{lllll}
\beta_{11}= 1/12 & \beta_{12}=1/12 & \beta_{13}= 2/15 & \beta_{14}= 16/315 & \beta_{15}=16/315  \\
\beta_{21}= 1/12 & \beta_{22}=1/4 & \beta_{23}= 1/5 & \beta_{24}= 4/105 & \beta_{25}=68/315 \\
\beta_{31}= 2/15 & \beta_{32}=1/5 & \beta_{33}= 4/15 & \beta_{34}= 16/315  & \beta_{35}=16/105 \\
\beta_{41}= 16/315 & \beta_{42}= 4/105 & \beta_{43}= 16/315 & \beta_{44}=64/945 & \beta_{45}= 64/2835 \\
\beta_{51}=332/105  & \beta_{52}=1852/105 & \beta_{53}= 592/63 & \beta_{54}=704/315 & \beta_{55}=704/2835.
\end{array}
$$

\end{document}